\pdfoutput=1 
\documentclass[%
 reprint,
 amsmath,amssymb,
aps,
pra,
]{revtex4-2}

\usepackage{graphicx}
\usepackage{dcolumn}
\usepackage{bm}
\usepackage{float}
\usepackage{siunitx}
\usepackage[hidelinks]{hyperref}

\usepackage{silence}
\usepackage{braket}

\usepackage[
justification = Justified 
]{caption}
\usepackage{subcaption}

\usepackage[capitalise]{cleveref}

\usepackage{xcolor}

\usepackage{color}

\usepackage{outlines}

\newcommand{\vek}[1]{\mathbf{#1}} 

\newcommand{\Tr}{\text{Tr}}

\newcommand{\dddt}[1]{\frac{d #1}{dt}}

\newcommand{\T}{T}

\newcommand{\V}[1]{\vek{V}_{#1 #1}}

\newcommand{\Vc}[1]{\vek{V}_{#1 #1}^c}

\newcommand{\phiC}{\phi^c}

\newcommand{\E}[1]{\text{E}\left[#1\right]}

\newcommand{\Real}{\text{Re}}

\newcommand{\fhat}[1]{\hat{#1}}

\newcommand{\fhatdag}[1]{\fhat{#1}^{\dagger}}

\newcommand{\hatvek}[1]{\hat{\vek{#1}}}
\newcommand{\fhatb}[1]{\hat{\mathbf{#1}}}

\newcommand{\Ham}{\fhat{H}}

\newcommand{\dmatrix}{\fhat{\rho}}

\newcommand{\LRWA}{\mathcal{L}_{\text{RWA}}}

\newcommand{\LnonRWA}{\mathcal{L}_{\text{nonRWA}}}

\newcommand{\Lenv}{\mathcal{L}_{\text{env}}}

\newcommand{\drivefield}{\epsilon}

\begin{document}


\title{Mechanical cooling and squeezing using optimal control}
\author{Frederik Werner Isaksen and Ulrik Lund Andersen}
\email{frwis@dtu.dk}
\email{ulrik.andersen@fysik.dtu.dk}
\affiliation{Center for Macroscopic Quantum States (bigQ), Department of Physics, Technical University of Denmark, 2800 Kgs. Lyngby, Denmark}


\begin{abstract}
A mechanical system can be optimally controlled through continuous measurements of its position followed by feedback. We revisit the complete formalism for predicting the performance of such as system without invoking the standard rotating wave approximations and the adiabatic approximation. Using this formalism we deduce both the conditional and unconditional state of a mechanical oscillator using the optimal control and feedback that leads to mechanical cooling and mechanical squeezing. We find large discrepancies between the exact solutions and the approximate solutions stressing the importance of using the complete model. We also highlight the importance of distinguishing between the conditional and unconditional state by demonstrating that these two cannot coincide in a typical control scheme, even with infinite feedback strength.
\end{abstract}

\date{\today}

\maketitle


\section{Introduction}

Fueled by the dramatic progress in developing high-quality nano- and micromechanical oscillators, there has recently been a surge of interest in controlling the motion of such oscillators at the quantum level for testing fundamental physics and for realizing novel quantum technologies\cite{Aspelmeyer2014,Barzanjeh2022}. A promising strategy for the optimal quantum control of a mechanical oscillator is by monitoring its motional dynamics through optimized measurements and subsequently use this information to drive the oscillator into a certain target state \cite{Doherty1999,Doherty2012}. Such a strategy has for example by now been used to prepare a mechanical oscillator near its quantum mechanical ground state\cite{Delic2020,Magrini2021,Rossi2018,Chan2011,Tebbenjohanns2021}. In addition to these experimental endeavours, there exists a vast theoretical literature on preparing mechanical oscillators in various quantum states via measurement-based feedback control including the ground state, the squeezed state \cite{Clerk2008,Meng2020} and more exotic states \cite{Hofer2015,Hoff2016}. 

The formalism of optimal feedback control of continuously measured quantum systems has been originally developed by Belavkin \cite{Belavkin1983,Belavkin1988,Belavkin1999} and later refined by Wiseman \& Milburn \cite{Wiseman2010}. It includes a complete quantum mechanical description of the control system and is often formulated in terms of a master equation for the density matrix representing the system. The formalism includes a real-time estimation algorithm that provides the optimal information about the measured state conditioned on previous measurements, and finally produces a conditional state which can be subsequently used to drive the mechanical oscillator into the desired state via feedback\cite{Doherty1999,Clerk2008,Wiseman2010,Doherty2012,Hamerly2013,Hofer2015,Yamamoto2017,Wieczorek2015,Setter2018,Rossi2019,Magrini2021,DiGiovanni2021}.

When using the master equation formalism to simulate an optomechanical system, several different approximations are often invoked. Possibly the two most important approximations are the rotating wave approximation (RWA) and the adiabatic approximation. The RWA can be applied when the dynamics of the mechanical system is much faster than all interactions with the environment and the measurement, while the adiabatic approximation is valid if the oscillator dynamics can be adiabatically followed by the probing system. While for some systems these two approximations can be taken, for others, however, they are not valid. As an example, the complete model predicts the formation of squeezed mechanical states via optimal feedback control while an approximative model based on the RWA of the interaction with the measurement apparatus cannot predict its appearance \cite{Meng2020}.

Moreover, it is important to distinguish the mechanical state that is inferred from the measurement record - known as the conditional state - and the mechanical state that is actually produced through active feedback control - known as the unconditional state. In much of the literature, these two states are often taken to be identical assuming that the feedback control can be done without any noise penalty. This is however a very crude assumption as decoherence of the mechanics plays an important role during feedback, rendering the unconditional state in a state that is more noisy than the conditional state.  

In this work we revisit the formalism for optimal feedback control using the master equation framework without using the RWA and the adiabatic approximation, and we apply the formalism to deduce the conditional and unconditional states of the mechanical oscillator. Our control parameters will be optimised for driving the mechanical oscillator into either a ground state or a squeezed state. We find, for example, that if the RWA of the interaction with the environment is applied, the optimal residual phononic occupancy when preparing the oscillator near its ground state is overestimated while when applying the adiabatic approximation, the squeezing degree is underestimated in certain regimes. Moreover, we show that the optimally prepared conditional and unconditional states are different, even if infinitely strong feedback is available. We discuss the consequences thereof, including for example how this changes the optimal measurement quadrature.

The rest of the paper is organized in three sections: In Sec.~\ref{sec:model} we present the model for optimal feedback control using the master equation framework while in Sec.~\ref{sec:results} we present and discuss the results of preparing mechanical ground states and squeezed states using our formalism with a special emphasis on the validity of the RWA and the adiabatic approximation. The experimental feasibility of the results is briefly discussed in Sec.~\ref{sec:feasibility}. The work is shortly summarized and concluded in Sec.~\ref{sec:conclusion}.


\section{The Physical Model}\label{sec:model}

\begin{figure}[h]
    \centering
    \includegraphics[width=\linewidth]{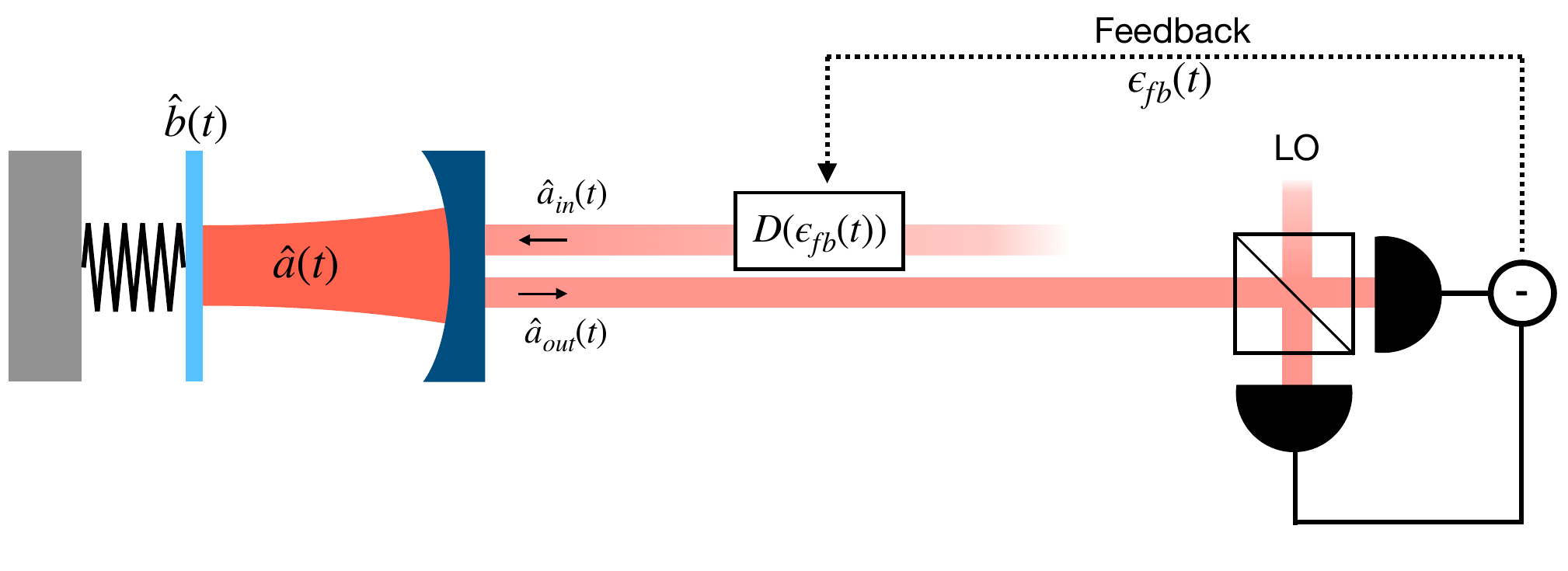}
    \caption{Sketch of the optomechanical setup; a cavity mode $\fhat{a}$ is interacting with a mechanical mode $\fhat{b}$. The output field is detected with a homodyne detector with a local oscillator (LO) with a phase corresponding to measurement of the output quadrature $\fhat{X}_{\text{out}}^{\theta}(t) = (\fhat{a}_{\text{out}}(t)e^{-i\theta}+\fhatdag{a}_{\text{out}}(t)e^{i\theta})/\sqrt{2}$. The homodyne detector has detection efficiency $\eta$. Based on the homodyne measurement signal, an optical feedback signal of complex amplitude $\drivefield_{\text{fb}}(t)$ is applied through a displacement operation $D(\drivefield_{\text{fb}}(t))$ $= \exp[\drivefield_{\text{fb}}(t) \fhatdag{a}_{\text{in}} -\drivefield_{\text{fb}}^*(t)\fhatdag{a}_{\text{in}}]$ on the input field.}
    \label{fig:model-skitse}
\end{figure}


We start by considering a standard model for the state of a mechanical oscillator which is conditioned on the continuous measurement of its position as illustrated in Fig.~\ref{fig:model-skitse}: A cavity mode with annihilation operator $\hat{a}$ and (angular) frequency $\omega_c$ is interacting through radiation-pressure forces with a mechanical oscillator with annihilation operator $\hat{b}$ and frequency $\Omega_m$. An input field $\fhat{a}_{\text{in}}$ of frequency $\omega_L$ is injected into the cavity with rate  $\kappa$. This input coherently drives the cavity field with frequency $\omega_L$ and time-dependent complex amplitude $\drivefield(t) =  \drivefield_{\text{probe}} +\drivefield_{\text{fb}}(t)$. Here, the constant term $\drivefield_{\text{probe}}$ is applied to enhance the optomechanical coupling and measurement strength, while the time-dependent term $\drivefield_{\text{fb}}(t)$ is a feedback induced control field. It is convenient to divide the latter contribution into real and imaginary parts, $\drivefield_{\text{fb}}(t) = (x_{\text{fb}} + i y_{\text{fb}})/\sqrt{2}$.
Working in a displaced frame rotating at the cavity field frequency $\omega_c$ and using a linearized approximation of the optomechanical radiation-pressure interaction, the full Hamiltonian of the system can be written as 
\begin{equation}\label{eq:hamiltonian}
\begin{split}
	\fhat{H} =& \hbar\Delta\fhatdag{a}\fhat{a} + \hbar\Omega_m\fhatdag{b}\fhat{b}  + 2\hbar g \fhat{Q}\fhat{X} \\
	 &+ \hbar\sqrt{\kappa}\left[x_{\text{fb}}(t)\fhat{Y}-y_{\text{fb}}(t)\fhat{X}\right]
\end{split}
\end{equation}
where $g$ is the probe-enhanced optomechanical coupling rate, $\hbar$ is Planck's reduced constant, $\fhat{Q} = (\fhat{b} + \fhatdag{b})/\sqrt{2}$ is the dimensionless position operator of the mechanics, and $\fhat{X} = (\fhat{a}+\fhatdag{a})/\sqrt{2}$ and $\fhat{Y} = (\fhat{a}-\fhatdag{a})/i\sqrt{2}$ are the cavity field amplitude and phase quadrature representations, respectively. We also introduce the dimensionless momentum $\fhat{P} = (\fhat{b}-\fhatdag{b})/i\sqrt{2} 
$ for later use.
Finally, $\Delta$ is the effective detuning, a parameter controlled by the laser frequency $\omega_L$. Throughout this paper, $\omega_L$ is chosen such that $\Delta = 0$. The reader is referred to Appendix~\ref{sec:linearisation} for a detailed derivation of the Hamiltonian in equation \eqref{eq:hamiltonian} as well as expressions for $g$ and $\Delta$.

Having specified the Hamiltonian of the cavity optomechanical system, we are now in a position to discuss the dissipative dynamics of the system. This will be done by using a stochastic master equation of the density matrix $\dmatrix$ representing the system. 

Both the cavity mode and the mechanical mode are inevitably subjected to loss and decoherence, and in this paper, we assume all these mechanisms to be Markovian. The cavity mode decays with rate $\kappa$ due to coupling to the input field. The mechanical mode is coupled to a thermal reservoir of average phonon occupation number $\bar{n}$ with damping rate $\Gamma_m$. As a result of this coupling, the quality factor of the mechanical oscillator is $Q_m = \Omega_m/\Gamma_m$.

Finally, the output field, represented by the field operator $\hat{a}_{\text{out}}$, is measured with a homodyne detector which is able to measure an arbitrary quadrature of the field given by $\fhat{X}_{\text{out}}^{\theta}(t) = (\fhat{a}_{\text{out}}(t)e^{-i\theta}+\fhatdag{a}_{\text{out}}(t)e^{i\theta})/\sqrt{2}$ where $\theta$ is determined by the phase of the detector's local oscillator. We will henceforth refer to $\theta$ as the \textit{phase} or the \textit{measurement angle} of the homodyne detector. The output field is related to the input field as per the usual input-output relations $\fhat{a}_{\text{out}} = \fhat{a}_{\text{in}} - \sqrt{\kappa}\fhat{a}$~\cite{Gardiner1985}. 
We assume that the detection efficiency of the homodyne detector is $\eta$. 

The information obtained by the continuous homodyne measurement produces a \textit{conditional} density matrix $\dmatrix_c$ of the joint system containing the cavity and the mechanical mode. As is customary in the literature, conditional dynamics will be explicitly indicated by a subscript $c$, i.e. $\dmatrix_c$ is the conditional density matrix and $\braket{\hat{A}}_c = \Tr[\hat{A}\dmatrix_c]$ is the conditional expectation value of the operator $\fhat{A}$ w.r.t. $\dmatrix_c$. This conditioning, combined with the Hamiltonian evolution including loss and decoherence, can be modeled by the following stochastic master equation \cite{Gardiner2000,Hofer2015}:
\begin{equation}\label{eq:master-equation}
\begin{split}
	d\dmatrix_c = &-\frac{i}{\hbar}[\Ham,\dmatrix_c]dt + \Lenv\dmatrix_c dt \\
	&+ \kappa\mathcal{D}[\fhat{a}]\dmatrix_c dt + \sqrt{\eta\kappa}\mathcal{H}[\fhat{a}e^{-i\theta}]\dmatrix_c dW.
\end{split}
\end{equation}
The superoperator $\Lenv$ describing the mechanical interaction with the environment is either $\LRWA$ or $\LnonRWA$, given by
\begin{subequations}\label{eq:mech-loss-operators}
\begin{align}
	\LRWA\dmatrix &= \Gamma_m(\bar{n} +1)\mathcal{D}[\fhat{b}]\dmatrix + \Gamma_m\bar{n}\mathcal{D}[\fhat{b}^{\dagger}]\dmatrix, \\
	\LnonRWA\dmatrix &= -\frac{i\Gamma_m}{2}[\fhat{Q},\{\fhat{P},\dmatrix\}] - \Gamma_m(\bar{n} + 1/2)[\fhat{Q},[\fhat{Q},\dmatrix]], \label{eq:mech-loss-nonRWA}
\end{align}
\end{subequations}
where $[\cdot,\cdot]$, and $\{\cdot,\cdot\}$ denote the commutator and anti-commutator, respectively, and the superoperators $\mathcal{D}$ and $\mathcal{H}$ are defined as
\begin{align}
\mathcal{D}[\fhat{c}]\dmatrix &:= \fhat{c}\dmatrix \fhatdag{c} - \frac{1}{2}\left(\fhatdag{c}\fhat{c}\dmatrix + \rho \fhatdag{c}\fhat{c}\right), \\
\mathcal{H}[\fhat{c}]\dmatrix &:= (\fhat{c}-\Tr[\fhat{c}\dmatrix])\dmatrix + \dmatrix(\fhatdag{c}-\Tr[\fhatdag{c}\dmatrix]).
\end{align}

The individual terms in Eqs.~\eqref{eq:master-equation} and \eqref{eq:mech-loss-operators} deserve some comments: The term $\LRWA\dmatrix$ is the Markovian rotating wave approximation to the environmental interaction with the mechanics, and is the most often used in the literature when modeling optomechanics with a master equation approach. On the other hand, the term $\LnonRWA\dmatrix$, first introduced in \cite{Caldeira1983} does not assume the rotating wave approximation, but also does not in general preserve positivity of the density matrix since it is not on Lindblad form. We refer the reader to ref. \cite{Giovannetti2001} for a discussion of some of the inadequacies of this system-environment master equation including proposed alternative models amending it to be on Lindblad form (See also Appendix~\ref{sec:app-positiviy}).

The terms in the second line of Eq.~\eqref{eq:master-equation} account for the effect on the master equation of cavity dissipation and subsequent homodyne detection of the output field when the measurement angle is $\theta$ \cite{Hofer2015}. $dW = dW(t)$ is the Wiener increment, a stochastic normally distributed variable satisfying the properties  $dW(t')dt = 0$, $dW(t)dW(t') = \delta_{t,t'}dt$, and $\E{dW(t)} = 0$, where $\E{\cdot}$ denotes the (classical) expectation value. The measured photocurrent corresponding to the above conditional evolution is \cite{Hofer2015}
\begin{equation}
\begin{split}
I(t)dt &= \sqrt{\eta\kappa}\langle \fhat{a}e^{-i\theta}+\fhat{a}^{\dagger}e^{i\theta}\rangle_c dt + dW(t) \\
&= \sqrt{2\eta\kappa}\langle \fhat{X}\cos(\theta) + \fhat{Y}\sin(\theta) \rangle_c dt + dW(t).
\end{split}    
\end{equation}
We remark that the above photocurrent $I(t)$ through homodyne detection is in fact singular due to the infinitesimal step size of $dt$ and $dW(t)$ \cite{Wiseman1993,Zoller1997,BowenMilburnOptomechanics}. The detected charge $q(t)$ defined by $dq(t) = I(t)dt$ is however well-defined.

Using the master equation in Eq.~\eqref{eq:master-equation}, equations of motion for the system operators $\fhat{\mathbf{X}} = (\fhat{Q},\fhat{P},\fhat{X},\fhat{Y})$ can now be derived. Assuming that the Wigner function of the initial state is Gaussian (e.g. a thermal state), the system will stay Gaussian.  This follows from the fact that homodyne detection preserves Gaussian states, as does time evolution under a Hamiltonian that is a second order polynomial of creation and annihilation operators \cite{Weedbrook2012}. Under this assumption, the quantum state $\dmatrix_c = \dmatrix_c(t)$ is then fully characterised by the mean vector $\braket{\hatvek{X}}_c$ and the covariance matrix $\vek{V}_{\vek{X}\vek{X}}^c = (V_{Z_1Z_2}^c)_{\hat{Z}_1,\hat{Z}_2 \in \hatvek{X}} = \Real(\braket{ \fhat{\mathbf{X}}\fhat{\mathbf{X}}^T }_c- \braket{\fhatb{X}}_c \braket{\fhatb{X}^T}_c) $, the equations of motion of which may be derived using the formula $d\langle \fhat{O} \rangle_c = \Tr[\fhat{O}d\dmatrix_c]$. We find (See Appendix~\ref{app:equations-of-motion}):
 \begin{subequations}\label{eq:conditional-state-space}
 \begin{align}
 	\begin{split}
    d\braket{\hatvek{X}}_c =& \left(\vek{A}\braket{\hatvek{X}}_c+\vek{B}\vek{u}\right)dt  \\
    &+ (\Vc{\vek{X}}\vek{C}^T + \vek{\Gamma}^T)dW
    \end{split} \label{eq:xc}
    \\
    \begin{split}
        \dddt{\Vc{\vek{X}}} =& \vek{A}\Vc{\vek{X}}+\Vc{\vek{X}}\vek{A}^T + \vek{D} 
        \\
         &- (\Vc{\vek{X}}\vek{C}^T + \vek{\Gamma}^T)(\vek{C}\Vc{\vek{X}} + \vek{\Gamma})
    \end{split} \label{eq:vc-matrix}
 \end{align} 
 \end{subequations}
Here, $\vek{u} = \vek{u}(t) = [x_{\text{fb}}(t)\;\;y_{\text{fb}}(t)]^T$ is a time-dependent vector describing the feedback of the system. The matrices $\vek{A}$ and $\vek{D}$ depend on which of the two dissipation models in Eq.~\ref{eq:mech-loss-operators} is used:
\begin{subequations}
\begin{align}
\mathbf{A}_{\text{RWA}} &= \begin{bmatrix}
-\Gamma_m/2 & \Omega_m & 0 & 0 \\
-\Omega_m & -\Gamma_m/2 & -2g & 0 \\
0 & 0 & -\kappa/2 & 0 \\
-2g & 0 & 0 & -\kappa/2
\end{bmatrix} \label{eq:a-matrix-rwa} \\
\mathbf{D}_{\text{RWA}} &= \begin{bmatrix}
\Gamma_m(\bar{n} + 1/2) & 0 & 0 & 0 \\
0 & \Gamma_m(\bar{n} + 1/2) & 0 &0 \\
0 & 0 & \kappa/2 & 0 \\
0 & 0 & 0 & \kappa/2
\end{bmatrix} \label{eq:d-matrix-rwa}
\end{align}
\end{subequations}

\begin{subequations}
\begin{align}
\mathbf{A}_{\text{nonRWA}} &= \begin{bmatrix}
0 & \Omega_m & 0 & 0 \\
-\Omega_m & -\Gamma_m & -2g & 0 \\
0 & 0 & -\kappa/2 & 0 \\
-2g & 0 & 0 & -\kappa/2
\end{bmatrix} \label{eq:a-matrix-nonrwa} \\
\mathbf{D}_{\text{nonRWA}}&= \begin{bmatrix}
0 & 0 & 0 & 0 \\
0 & 2\Gamma_m(\bar{n} + 1/2) & 0 &0 \\
0 & 0 & \kappa/2 & 0 \\
0 & 0 & 0 & \kappa/2
\end{bmatrix} \label{eq:d-matrix-nonrwa}
\end{align}
\end{subequations}
On the other hand, $\vek{B}$, $\vek{C}$ and $\vek{\Gamma}$ are the same for both models:

\begin{subequations}
\begin{align}
	\vek{B} &= \begin{bmatrix}
         0 & 0 \\
        0 & 0 \\
         \sqrt{\kappa} & 0 \\
         0 & \sqrt{\kappa}
    \end{bmatrix} \\
    \vek{C} & = \sqrt{2\eta\kappa}\begin{bmatrix}
        0 & 0 & \cos(\theta) & \sin(\theta)
    \end{bmatrix} \\
    \vek{\Gamma} &= \sqrt{\frac{\eta \kappa}{2}}\begin{bmatrix}
        0 & 0 & -\cos(\theta) & -\sin(\theta)
    \end{bmatrix}
\end{align}
\end{subequations}
We note that Eqs.~\eqref{eq:conditional-state-space} are formally equivalent to a Kalman Filter \cite{Wiseman2010}. For the rest of this paper, we will mainly be concerned with the steady-state dynamics of $\Vc{\vek{X}}$, i.e. the solution when $\dddt{}\Vc{\vek{X}} = 0$. Equation \eqref{eq:vc-matrix} then reduces to an algebraic equation in $\Vc{\vek{X}}$ known as the algebraic Riccati equation. Onwards, $\Vc{\vek{X}}$ will therefore almost exclusively refer to the steady-state solution.

Before proceeding, we will make a few remarks on the adiabatic approximation. This approximation is regularly made in the so-called bad cavity regime where $\kappa/\Omega_m \gg 1$. In this regime, photons that enter the cavity will exit after some time that is much shorter than the mechanical period. Thus from the perspective of a photon, the mechanical oscillator essentially does not move during its short stay inside the cavity. The adiabatic approximation is assuming that the cavity field immediately reaches a dynamical steady-state solution that depends on the mechanical motion. In Ref.~\cite{Meng2020}, this approximation is used to find analytical expressions of the  conditional covariance matrix of the mechanics under continuous measurement of the phase quadrature of the light using the nonRWA model (albeit using a different theoretical framework involving quantum Langevin equations and Wiener filtering instead of conditional master equations as employed here). We will later compare the results found in Ref.~\cite{Meng2020} with our results which do not invoke the adiabatic approximation.

\begin{figure}[H]
	\centering
	\includegraphics[trim = {0 1.3cm 0 1.8cm},clip,width = \linewidth]{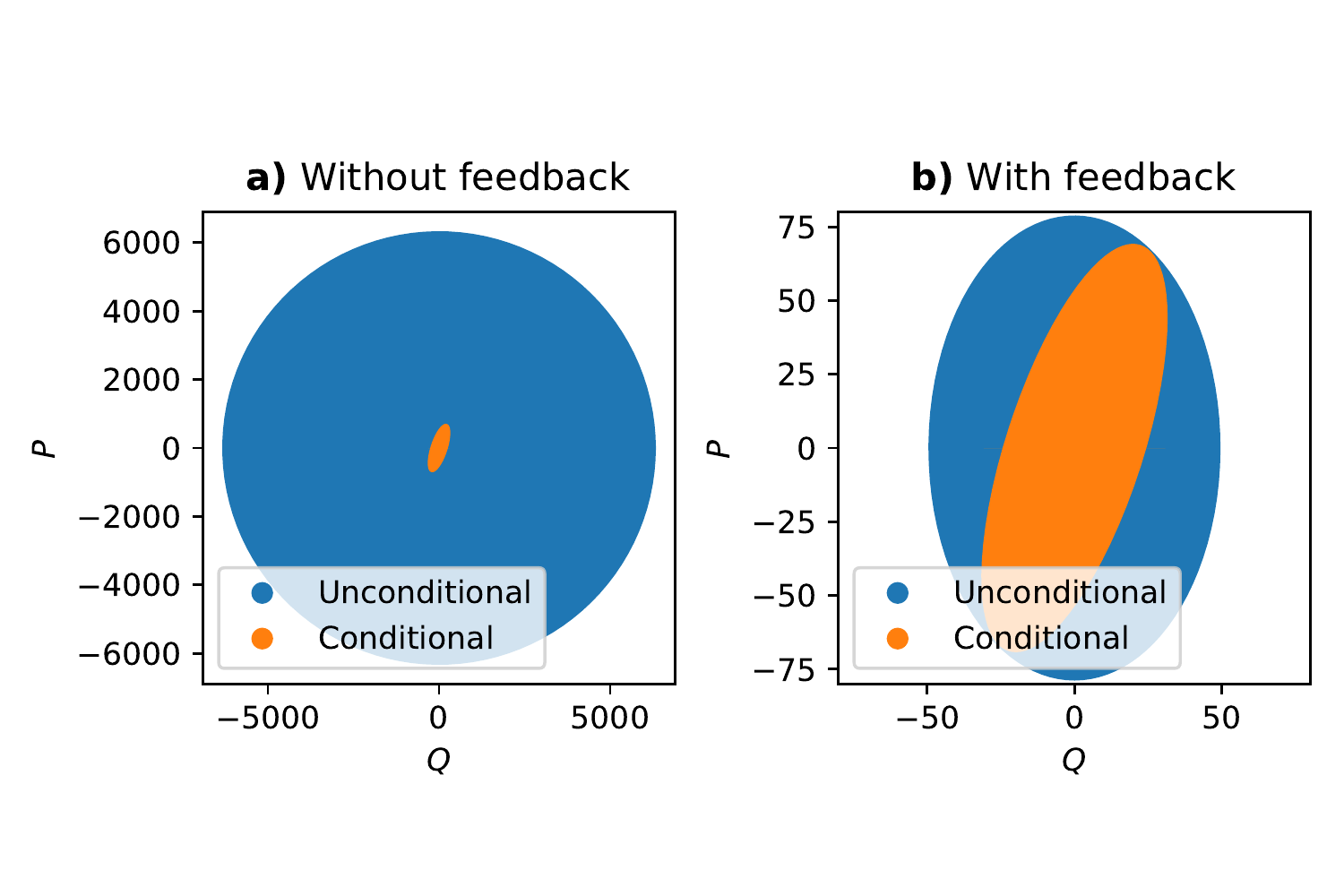}
	\caption{The mechanical covariance matrix in steady-state depicted as an uncertainty ellipsis (1 standard deviation) in $(Q,P)$ phase space. Without feedback (Fig.~\ref{fig:conditional-vs-uncondtitional-sketch}a), the unconditional state is much larger compared to when feedback is applied (Fig.~\ref{fig:conditional-vs-uncondtitional-sketch}b), but the conditional state is unchanged. In Fig.~\ref{fig:conditional-vs-uncondtitional-sketch}a, the area of the conditional uncertainty ellipse have been artificially scaled up by a factor of 100 to make it more visible. Plotted for $\Omega_m = \SI{e6}{Hz}$, $Q_m = 10^4$, $g = \SI{e5}{Hz}$, $\kappa = \SI{e8}{Hz}$, $T = \SI{300}{K}$, $\eta = 1$, $q = \num{e-5}$, $p = \num{e3}$, using the nonRWA model and with feedback that minimizes the phonon number.}
	\label{fig:conditional-vs-uncondtitional-sketch}
\end{figure}

Up to this point, we have considered the conditional state of the system, $\dmatrix_c$. However, of equal importance is the  \textit{unconditional} state $\dmatrix := \E{\dmatrix_c}$. One computes expectation values from this state using the definition $\braket{\cdot} := \Tr[\cdot\dmatrix] = \E{\braket{\cdot}_c}$, where the final equality is straightforward to prove. The qualitative difference between $\dmatrix$ and $\dmatrix_c$ is as follows: $\dmatrix_c$ depends on (i.e. is conditioned on) the outcome of the homodyne measurements up to time $t$. As these measurement outcomes are probabilistic in nature, the trajectory of the state will follow a random path in phase space (customarily called \textit{quantum trajectory}). On the other hand, the unconditional state is the average over all possible trajectories from the beginning of the experiment up to time $t$. For a feedback $\vek{u}$ equal to zero, this basically corresponds to the state of system in which the measurement outcomes are ignored. Mathematically, this is equivalent to removing the term proportional to $dW$ in Eq.~\eqref{eq:master-equation}, which means that the $dW$ term in Eq.~\eqref{eq:xc} and the term $- (\Vc{\vek{X}}\vek{C}^T + \vek{\Gamma}^T)(\vek{C}\Vc{\vek{X}} + \vek{\Gamma})$ in Eq.~\eqref{eq:vc-matrix} will vanish. The result is that the mean vector is identically zero for all $t$, but that the covariance is much larger due to large contributions from both thermal noise and backaction noise. The difference between the conditional and unconditional states of the mechanical mode (with the cavity mode traced out) is illustrated in phase space in Fig.~\ref{fig:conditional-vs-uncondtitional-sketch}a.
Note that \textit{both} $\dmatrix$ and $\dmatrix_c$ are depicted as having zero mean in Fig.~\ref{fig:conditional-vs-uncondtitional-sketch} for easier comparison, even though $\dmatrix_c$ in general will be displaced from the origin by some amount conditional on the measurement record. 
 
To prepare the unconditional state into a state that resembles the conditional state, we need to actively control the motion of the mechanical oscillator based on the homodyne measurement outcomes. As illustrated in Fig.~\ref{fig:conditional-vs-uncondtitional-sketch}, the results of the homodyne measurements are fed back onto the optomechanical system, thereby driving it into a low-entropy unconditional state as described by the Hamiltonian in Eq.~(\ref{eq:hamiltonian}). A phase space representation of the unconditional state with feedback is shown in Fig.~\ref{fig:conditional-vs-uncondtitional-sketch}b where it is compared to the conditional state for a particular representative example. The resulting unconditional state depends critically on the estimation and feedback strategy which in turn depends on the state one aims to prepare. In the following we consider the optimal feedback strategies applied for the preparation of an unconditional state with a minimal phonon number (corresponding to mechanical cooling) as well as the optimal strategy for minimizing the variance of one of the mechanical quadratures (corresponding to mechanical squeezing).



\subsection{Optimal control formalism}

In this section we consider the optimal control schemes for mechanical cooling and squeezing. We use the framework of linear-quadratic-Gaussian (LQG) optimal control which is tailored to minimize a quadratic \textit{cost function} \cite{Wiseman2010}:
\begin{equation}\label{eq:cost-function-j}
	j = \int_{t_0}^{t_1} \E{\Tr\left\{\dmatrix_c(t)h(\hatvek{X},\mathbf{u}(t),t)\right\}}dt.
\end{equation}
Here, $t_0$ and $t_1$ denote the start and end time of the run, respectively, and
\begin{equation}\label{eq:cost-function-h}
\begin{split}
	h(\hatvek{X},\mathbf{u}(t),t) =& \hatvek{X}^T\mathbf{P}\hatvek{X}+\mathbf{u}(t)^T\mathbf{Q}\mathbf{u}(t)
	\\
	&+ 2\delta(t-t_1)\hatvek{X}^T\mathbf{P}_1\hatvek{X},
\end{split}
\end{equation}
where $\mathbf{P}$ , $\mathbf{P}_1$ and $\mathbf{Q}$ are matrices that are specified according to the problem one wants to solve, for example minimizing the mechanical phonon occupancy or minimizing a specific quadrature variance of the mechanical oscillator.
The value of $\mathbf{P}_1$ is associated with a terminal cost at time $t_1$. However, since we are only interested in the dynamics at steady state, the value of $\mathbf{P}_1$ is irrelevant. This is commonly referred to as an \textit{asymptotic}  or \textit{infinite horizon} LQG problem. 

According to standard optimal control theory, the optimal feedback is then given by
\begin{equation}
	\mathbf{u}(t) = -  \vek{K(t)}\langle \hatvek{X} \rangle_c(t),
\end{equation}
where $\mathbf{K}(t)=\mathbf{Q}^{-1}\mathbf{B}^T\mathbf{Y}(t)$ is the Kalman gain and $\mathbf{Y}(t)$ is a symmetric, positive semi-definite matrix satisfying the differential equation
\begin{equation}\label{eq:kalman-gain-diffeq}
	-\frac{d\mathbf{Y}(t)}{dt} =  \mathbf{A}^T\mathbf{Y}(t) + \mathbf{Y}(t)\vek{A} + \mathbf{P} - \vek{Y}(t)\vek{B}\mathbf{Q}^{-1}\mathbf{B}^T\mathbf{Y}(t)
\end{equation}
with the terminal condition $\mathbf{Y}(t_1) = \mathbf{P}_1$.

For asymptotic LQG-problems, $t_1-t_0$ is very large compared with all other rates of our system. Within this time period, it can be shown that there exists a steady-state solution $\vek{Y}$ to Eq.~\eqref{eq:kalman-gain-diffeq} (i.e. with $-\dddt{}\vek{Y} = 0$) assuming certain stabilization conditions \cite{Wiseman2010}. Given these conditions and with this choice of feedback, the steady-state \textit{unconditional variance}  $\V{\vek{X}} = \Real(\braket{ \fhat{\mathbf{X}}\fhat{\mathbf{X}}^T } )- \braket{\fhatb{X}} \braket{\fhatb{X}^T})$ of the system is given by the relation $\V{\vek{X}}= \Vc{\vek{X}}+ \V{\vek{X}}^E$, where $\V{\vek{X}}^E = \E{\braket{\hatvek{X}}_c\braket{\hatvek{X}}_c^T}$ is an excess noise contribution stemming from imperfect feedback. This excess noise variance satisfies the equation
\begin{equation}\label{eq:unconditonal-cov-steady-state}
	\vek{N}\V{\vek{X}}^E+\V{\vek{X}}^E\vek{N}^T + \vek{F}^T\vek{F} = 0,
\end{equation}
where $\mathbf{N} = \mathbf{A} - \mathbf{B}\mathbf{K}$ and $\mathbf{F}= \mathbf{C}\Vc{\vek{X}}+\mathbf{\Gamma}$.
The expressions Eq.~\eqref{eq:kalman-gain-diffeq} and Eq.~\eqref{eq:unconditonal-cov-steady-state}, together with the conditional dynamics in Eq.~\eqref{eq:conditional-state-space} gives the complete unconditional steady-state dynamics of the system, and in particular the steady-state unconditional variances, $\V{\vek{X}}$.

\subsection{Optimal control for mechanical cooling}
Having outlined the overall strategy of optimal control, we will now consider two specific examples of optimal control associated with mechanical cooling and mechanical squeezing. 

A minimization of the phonon number is obtained by setting $\mathbf{P} = \text{diag}(p \Omega_m,p \Omega_m,0,0)$ and $\mathbf{Q} = \text{diag}(q,q)$ where $p$ and $q$ are arbitrary dimensionless parameters. 
Note that $\mathbf{Q}$ specifies the cost associated with the feedback scheme, and that the fraction $p/q$ is a measure of the feedback power for minimizing the oscillator energy. By using these particular matrices for $\mathbf{P}$ and $\mathbf{Q}$ as well as the formalism in Eqs.~\eqref{eq:kalman-gain-diffeq} and \eqref{eq:unconditonal-cov-steady-state}, we find $\V{\vek{X}}$, and subsequently the minimized phonon number $n = \left(V_{QQ}+ V_{PP} -1 \right)/2$.

\subsection{Optimal control for mechanical squeezing}
The above procedure can also be used for minimizing the mechanical variance. Specifically, in order to minimize $V_{Q_{\nu}Q_{\nu}}$, that is, the variance along $\hat{Q}_{\nu} = \cos(\nu)\hat{Q} +\sin(\nu)\hat{P}$ for some angle $\nu$, it is straightforward to show that the appropriate cost function is given as in Eqs.~\eqref{eq:cost-function-j} and \eqref{eq:cost-function-h}, but with
\begin{equation}\label{eq:minimum-variance-pmatrix}
	\vek{P} = p\Omega_m\begin{bmatrix}
		\cos^2(\nu) & \cos(\nu)\sin(\nu) & 0 & 0 \\
		\cos(\nu)\sin(\nu) & \sin^2(\nu) & 0 & 0 \\
		0 & 0 & 0 & 0 \\
		0 & 0 & 0 & 0		
	\end{bmatrix},
\end{equation}
while $\vek{Q}$ is identical to the case of mechanical cooling. The minimum variance is then $\min_{\nu}(V_{Q_{\nu}Q_{\nu}}) =: V_{Q_{\phi}Q_{\phi}}$ and $\phi$, referred to as the squeezing angle, is the optimal choice of $\nu$. 
Note that we will also investigate the minimum conditional variance $V_{Q_{\phiC}Q_{\phiC}}^c$, which in general has a different squeezing angle $\phiC$.

\subsection{Asymptotic feedback}\label{sec:asymptotic-feedback}
 In general, analytical solutions to the steady-state equations governing the conditional and unconditional covariance matrices do exists, but are unwieldy large and thus too impractical. However, simple expressions for the excess covariance matrix elements in $\V{\vek{X}}$ can be derived in the important limit where $p/q \rightarrow \infty$, i.e. when the feedback cost is negligible. In this section we will be presenting such solutions for the two cases discussed in the previous sections. We will be assuming the generally valid conditions: $g,\kappa,\Omega_m > 0$, $Q_m > 1/2$  
 and $V^c_{Q,X_{\theta}}  \neq 0$. This last condition is naturally obeyed due to the opto-mechanical coupling. Detailed derivations are found in Appendix~\ref{app:asymptotic}.

We first consider the feedback strategy that minimizes the phonon number $n$. Applying the RWA, we find the steady-state Kalman gain $\vek{K} = \vek{Q}^{-1}\vek{B}^{\T}\vek{Y}$ to leading order in $p/q$:
\begin{equation}\label{eq:asymp-kalman-gain}
	\vek{K}^{\T} = \begin{bmatrix}
		\frac{1}{2}(Q_m^{-1} - \sqrt{4+Q_m^{-2}})\sqrt{\frac{p}{q}} + O\left[\left(\frac{p}{q}\right)^{1/4}\right] &  0 \\
		\frac{1}{2}\sqrt{\frac{p}{q}} +O\left[\left(\frac{p}{q}\right)^{1/4}\right] & 0 \\
		2\sqrt{\frac{g}{\sqrt{\kappa\Omega_m}}}\left(\frac{p}{q}\right)^{1/4} + O(1) & 0 \\
		0 & 0 \\
	\end{bmatrix}
\end{equation}
We note that $\vek{K}$ grows unboundedly with $p/q$, implying that infinite feedback strength is required in that particular limit. We also note that the entries of the fourth row and column are all zero, implying that information about the phase quadrature $\braket{\hat{Y}}$ will not be used in the Kalman gain, and that the feedback will not be applied to the phase input quadrature, i.e. $y_{\text{fb}}(t) = 0$.

Inserting this expression into Eq.~\eqref{eq:unconditonal-cov-steady-state}, and taking the limit $p/q \rightarrow \infty$, we find the excess covariance matrix elements relevant for the phonon numbers to
\begin{align}
V_{QQ}^E &\rightarrow \frac{2}{\sqrt{4+Q_m^{-2}}}\frac{\eta\kappa}{\Omega_m} 
(V_{Q X_{\theta}}^{c})^2 \\
V_{PP}^E &\rightarrow \frac{\left(2+Q_m^{-2}-Q_m^{-1}\sqrt{4+Q_m^{-2}}\right)}{\sqrt{4+Q_m^{-2}}}\frac{\eta\kappa}{\Omega_m}
(V_{Q X_{\theta}}^{c})^2,
\end{align}
for $p/q \rightarrow \infty$.

For the nonRWA model, similar calculations yield
\begin{subequations}\label{eq:excess-noise-cooling}
\begin{align}
V_{QQ}^E &\rightarrow \frac{\eta\kappa}{\Omega_m} 
(V_{Q X_{\theta}}^c)^2, \label{eq:excess-noise-cooling-qq}  \\
V_{PP}^E &\rightarrow \frac{\eta\kappa}{\Omega_m}(V_{Q X_{\theta}}^c)^2.
\label{eq:excess-noise-cooling-pp}
\end{align}
\end{subequations}
for $p/q \rightarrow \infty$. It is clear from these expressions that both for the RWA and nonRWA, the excess noise associated with the preparation of an unconditional state is quadratically proportional to the correlations between the position of the mechanics and the quadrature, $\fhat{X}_{\theta}$ of light.   

If we instead choose the feedback strategy that minimizes the minimum variance $V_{Q_{\nu}Q_{\nu}}$ (corresponding to the choice of $\vek{P}$ in Eq.~\eqref{eq:minimum-variance-pmatrix}), we find for the nonRWA case that:
\begin{subequations}\label{eq:excess-noise-squeezing}
\begin{align}
V_{Q_{\nu}Q_{\nu}}^E &\rightarrow \frac{\eta\kappa}{\Omega_m}(V_{Q X_{\theta}}^c)^2\begin{cases}
-2 \sin(2\nu),& \nu \in ]-\pi/2, 0[ \\
0,& \nu = [0,\pi/2]
\end{cases} \label{eq:excess-noise-squeezing-qq} \\
V_{P_{\nu}P_{\nu}}^E &\rightarrow \frac{\eta\kappa}{\Omega_m}(V_{Q X_{\theta}}^c)^2\begin{cases}
-\frac{\cos(4\nu) + 1}{\sin(2\nu)}, & \nu \in ]-\pi/2,0[ \\
\infty,& \nu = 0 \\
2\csc(2\nu), & \nu \in ]0,\pi/2[ \\
\infty, & \nu = \pi/2
\end{cases} \label{eq:excess-noise-squeezing-pp} \\
V_{Q_{\nu}P_{\nu}}^E &\rightarrow \frac{\eta\kappa}{\Omega_m}(V_{Q X_{\theta}}^c)^2\begin{cases}
-2\cos(2\nu),& \nu \in ]-\pi/2,0[ \\
-1,& \nu = 0, \\
0,&  \nu \in ]0,\pi/2[ \\
1,& \nu = \pi/2,
\end{cases}
\label{eq:excess-noise-squeezing-qp}
\end{align}
\end{subequations}
for $p/q \rightarrow \infty$. The above expressions are provided only for $\nu \in ]-\pi/2,\pi/2]$ since they are $\pi$-periodic. 

A number of important observations can be made from these results.
The Kalman filter and optimal control stratedy applied here with $p/q \rightarrow \infty$ yields the lowest possible unconditional phonon number or minimum variance, given that we are able to measure the output field through homodyne detection and apply feedback by displacing the input field as described. However, even in this optimal limit the unconditional dynamics do not match the conditional dynamics. The unconditional phonon number for mechanical cooling as well as the minimized variance for mechanical squeezing  will have an extra contribution quantified by the excess noise term in Eqs.~\eqref{eq:excess-noise-cooling} and Eqs.~\eqref{eq:excess-noise-squeezing}, respectively as also illustrated in Fig.~\ref{fig:conditional-vs-uncondtitional-sketch}b.  
This is in contrast to what appears to be a common belief in the literature, namely that the unconditional state may always approach the conditional state under feedback. 

\section{Results and discussion}\label{sec:results}

In this section we will be using the mathematical framework for optimal quantum control derived in the previous section for estimating the minimum phonon occupancy as well as the maximum amount of squeezing of the mechanical oscillator. We will both investigate the conditional as well as the unconditional state, and consider the effects of the rotating wave approximation and the adiabatic approximation. Note that the plots of the adiabatic approximation in Figs.~\ref{fig:n-vs-Omega_m-adiabadic-comparison}, \ref{fig:Vc_Qphi-vs-Omega_m-and-g-MB2020-comparison}, and \ref{fig:Asymp-Vqmin-theta_opt-phi-vs-omega-g-is-1e7} are based on the analytical expressions of the conditional covariance matrix found in Ref.~\cite{Meng2020}. Note also that all plots of unconditional phonon numbers and variances are in the zero feedback cost limit $p/q \rightarrow \infty$. Finally, to guard against potential unphysical results from the nonRWA-model as discussed in Sec.~\ref{sec:model}, we checked all numerical results in this paper to make sure that the positivity of the density matrix was always preserved, see Appendix~\ref{sec:app-positiviy}.



\subsection{Mechanical cooling}
\begin{figure}[t]
	\centering
    \includegraphics[width = \linewidth]{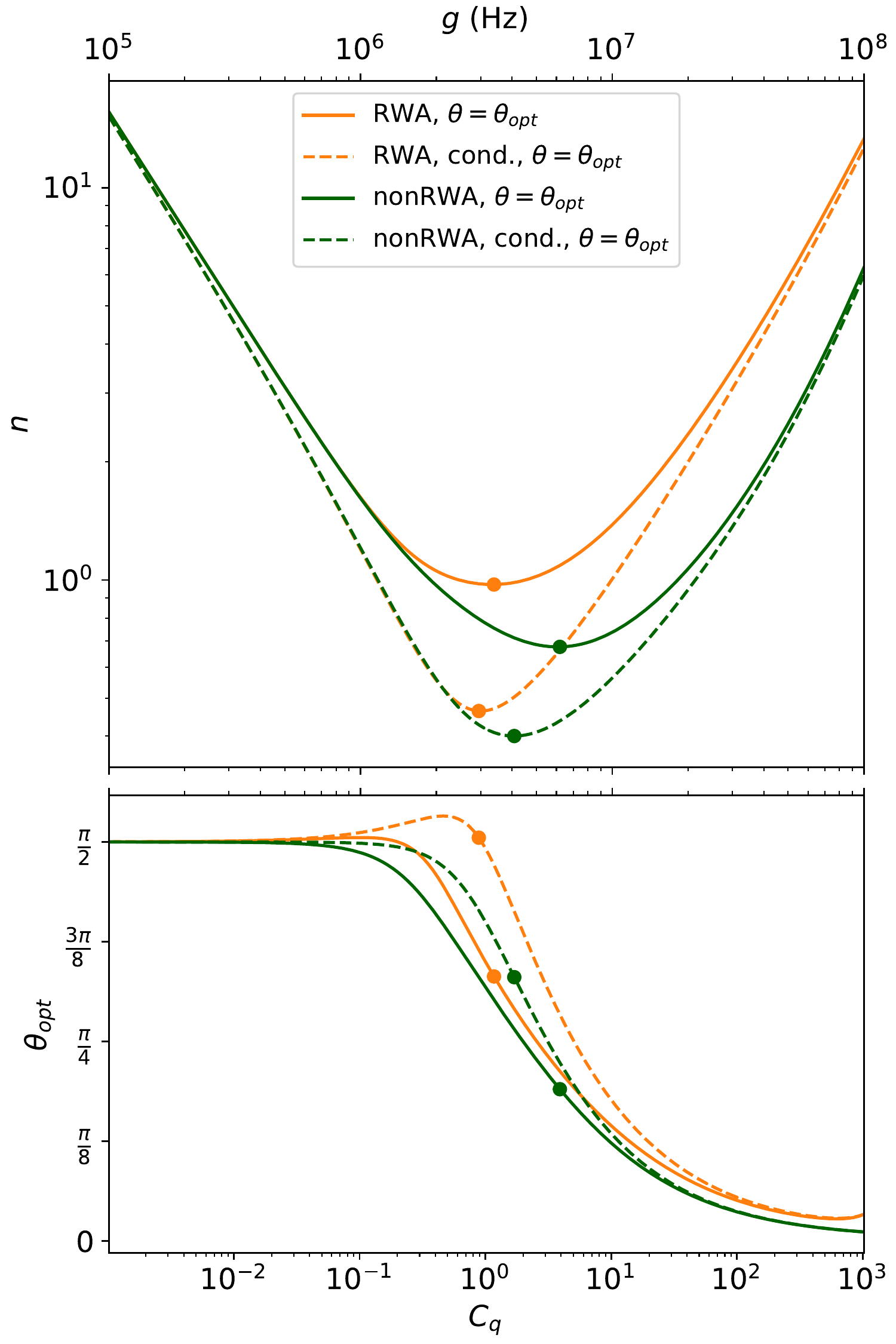}
	\caption{Top: Conditional (cond., dashed lines) and unconditional (solid lines) phonon numbers $n$ vs coupling strength $g$ for the RWA and nonRWA model for $\theta = \theta_{\text{opt}}$. The minimum values of $n$ as a function of $g$ are marked with filled-in circles. Parameters used for this plot is $\Omega_m = \SI{e6}{Hz}$, $\kappa = \SI{e8}{Hz}$, $Q_m = \num{e8}$, and $\eta = 1$. Bottom: corresponding values of $\theta_{\text{opt}}$ versus $g$. The values at which $n$ is minimal are marked with filled-in circles.}
	\label{fig:n-and-theta-opt-vs-g}
\end{figure}

In this section we will consider the effect of measurement induced cooling both on the conditional state where the state of the mechanical system is inferred by the measurements, and the unconditional state where the measured information is actively fed back onto the oscillator to drive it into a low entropy state. By using the formalism in the previous section, we plot in Fig.~\ref{fig:n-and-theta-opt-vs-g} the minimum mechanical phonon occupancy against the opto-mechanical coupling strength both in terms of the coupling parameter, $g$, and the quantum cooperativity, $C_q=4g^2/(\kappa\Gamma_m\bar{n})$ where $\bar{n}= \left\{\exp(\hbar\Omega_m/k_{\text{B}}T) - 1 \right\}^{-1}$. The environmental temperature $T$ is set to 300 K throughout the article. The plots in Fig.~\ref{fig:n-and-theta-opt-vs-g} have been numerically optimized (that is, the phonon number has been minimized) over the measurement angle $\theta$ of the local oscillator. These optimal angle $\theta_{\text{opt}}$ are also shown in Fig.~\ref{fig:n-and-theta-opt-vs-g}. 

It is clear that for low coupling strengths, the optimal measurement angle is $\pi/2$ as expected since in that case all the information about the mechanical oscillator is transferred to the phase quadrature of the probe field. However, for larger coupling strengths, the radiation pressure force creates correlations between the amplitude quadrature of light and the mechanical position, which can be advantageously used for cooling by rotating the phase angle away from $\pi/2$. For very strong coupling, the optimal angle $\theta_{\text{opt}}$ tends to $0$ which corresponds to an amplitude quadrature measurement. 

We also clearly see from Fig.~\ref{fig:n-and-theta-opt-vs-g} that by imposing the RWA between the oscillator and the environment, the phononic occupancy is in general underestimated, in particular in the strong coupling regime. For weak coupling, the measurement rate will be low, which means that several mechanical oscillations will be required to resolve its motion. As a result, detailed information about the position and momentum of the mechanical oscillator is being washed out, and thus a potential asymmetry in phase space of the mechanical state due to its coupling to the environment is not visible. In this regime, the RWA is therefore completely valid. However, when the measurement rate becomes large (that is, stronger coupling), the two mechanical quadratures can be better resolved, and thus a potential asymmetry imposed by the coupling to the environment becomes visible in the measurement. To capture these phase space correlations imposed by the environment, only the model without the RWA is valid, and as seen by Fig.~\ref{fig:n-and-theta-opt-vs-g} (comparing the RWA with nonRWA curves), the knowledge of these correlations indeed improves the cooling rate. The deviation between the RWA and nonRWA models naturally depends on the strength of the coupling of the mechanics to the environment (relative to the mechanical frequency) which is dictated by the thermalization rate, $\gamma_{th}=\Gamma_m(\bar{n}+1/2)$. In the following plots, we will compare the phonon number against the \textit{normalized} thermalization rate  $\tilde{\gamma}_{th} := \gamma_{th}/\Omega_m$. Note that for constant $Q_m$, this number is roughly inversely proportional to $\Omega_m$ as $\tilde{\gamma}_{th} = Q_m^{-1}(\bar{n}+1/2) \simeq Q_m^{-1}k_BT/(\hbar \Omega_m)$ for large $\bar{n}$.

It is clear from Fig.~\ref{fig:n-and-theta-opt-vs-g} that the phonon occupancy is minimized for a certain value of the cooperativity (measurement rate). This is attributed to the fact that a large measurement rate will make the measurement sharp and thus prepare the mechanical state in a squeezed state which inevitably adds phonons to the state. The difference between the conditional and unconditional state is also clearly evident from Fig.~\ref{fig:n-and-theta-opt-vs-g}, and as expected the conditional state has a lower entropy than the unconditional state. The feedback required for the generation of the unconditional state imposes a noise penalty due to decoherence of the mechanics during feedback. This extra excess noise is however small, and plays only a role for small phonon occupancy: When the cooperativity is large, the phonon occupancy becomes large and the extra feedback-induced excess noise is negligible.

\begin{figure}[ht] 
	\centering
    \includegraphics[width = \linewidth]{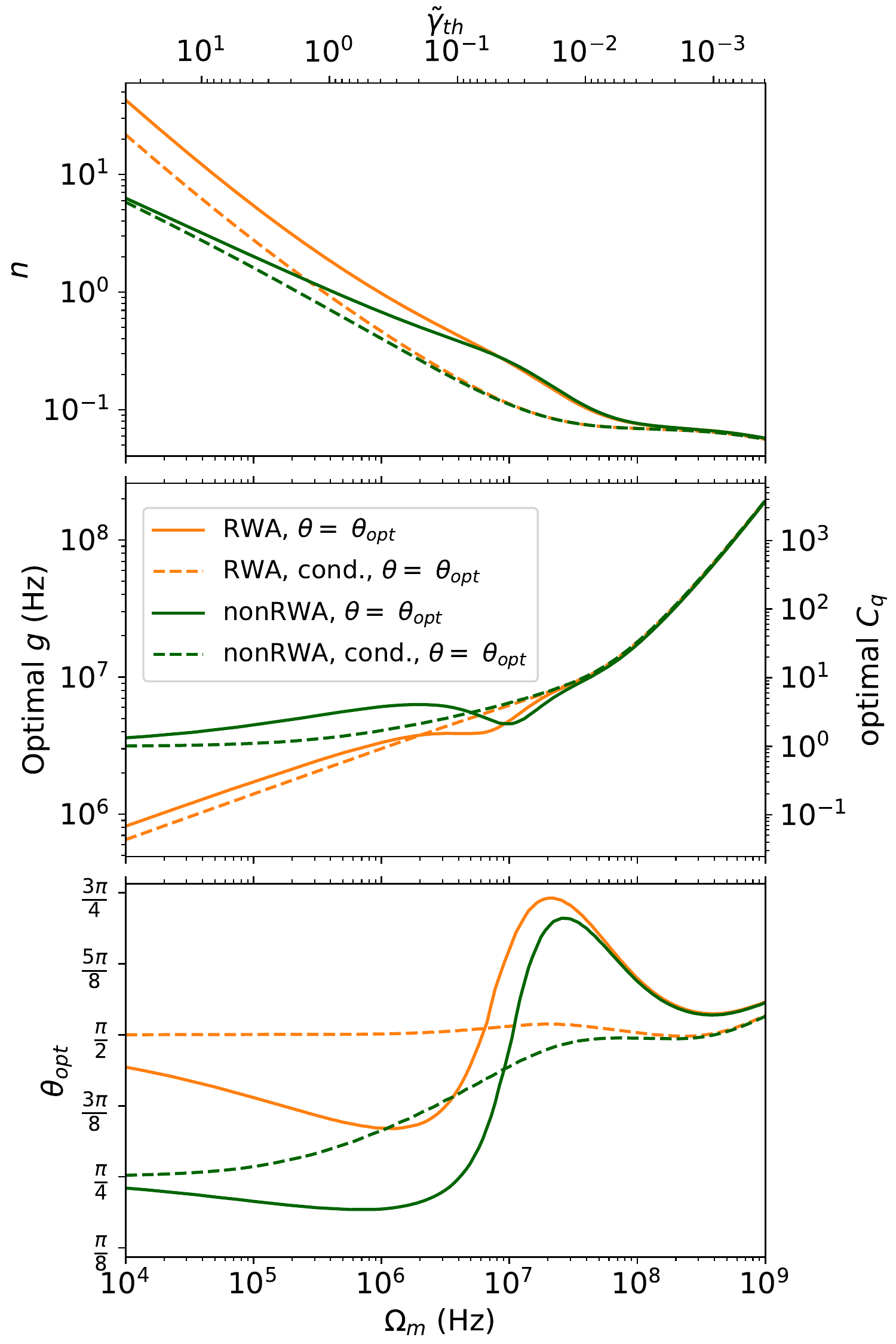}
	\caption{Plots of conditional (cond., dashed lines) and unconditional (solid lines) phonon numbers $n$ at the optimal coupling strength $g$ versus mechanical frequency $\Omega_m$ and normalized thermalization rate $\tilde{\gamma}_{th} \propto \Omega_m^{-1}$ for the RWA and nonRWA model for $\theta = \theta_{\text{opt}}$. Top: phonon number $n$ is plotted. Middle: Plots of the optimal values of $g$ and equivalent values of the optimal quantum cooperativity $C_q$. Bottom: Plots of the optimal value of the measurement angle, $\theta_{\text{opt}}$. Parameters used in this figure are $\kappa = \SI{e8}{Hz}$, $Q_m = 10^8$, and $\eta = 1$.}
	\label{fig:n-g-theta--vs--Omega_m}
\end{figure}

In Fig.~\ref{fig:n-g-theta--vs--Omega_m}, top, we show the minimized phonon occupancy against the normalized thermalization rate $\tilde{\gamma}_{th}$ in which we have optimized the values of the coupling strength, $g$ (or $C_q$) (shown in Fig.~\ref{fig:n-g-theta--vs--Omega_m}, middle) and the measurement angle, $\theta$ (shown in Fig.~\ref{fig:n-g-theta--vs--Omega_m}, bottom). The phonon occupancy is again illustrated both for the approximative model applying the RWA as well as the complete solution without relying on the RWA. As expected, we observe a large deviation between the two models when the thermalization rate is large (or equivalently when the mechanical frequency is low). This is caused by the establishment of mechanical quadrature correlations due to the strong environmental coupling which is neglected by the RWA model. We also again observe a large difference between the conditional and the unconditional states which is attributed to decoherence during feedback, and which is negligible for low decoherence rates. It is also interesting to note the large difference in the optimal measurement phases for the conditional and unconditional cases which is caused by the complex dynamics that the mechanics undergo during feedback.

\begin{figure}[ht] 
	\centering
    \includegraphics[width=\linewidth]{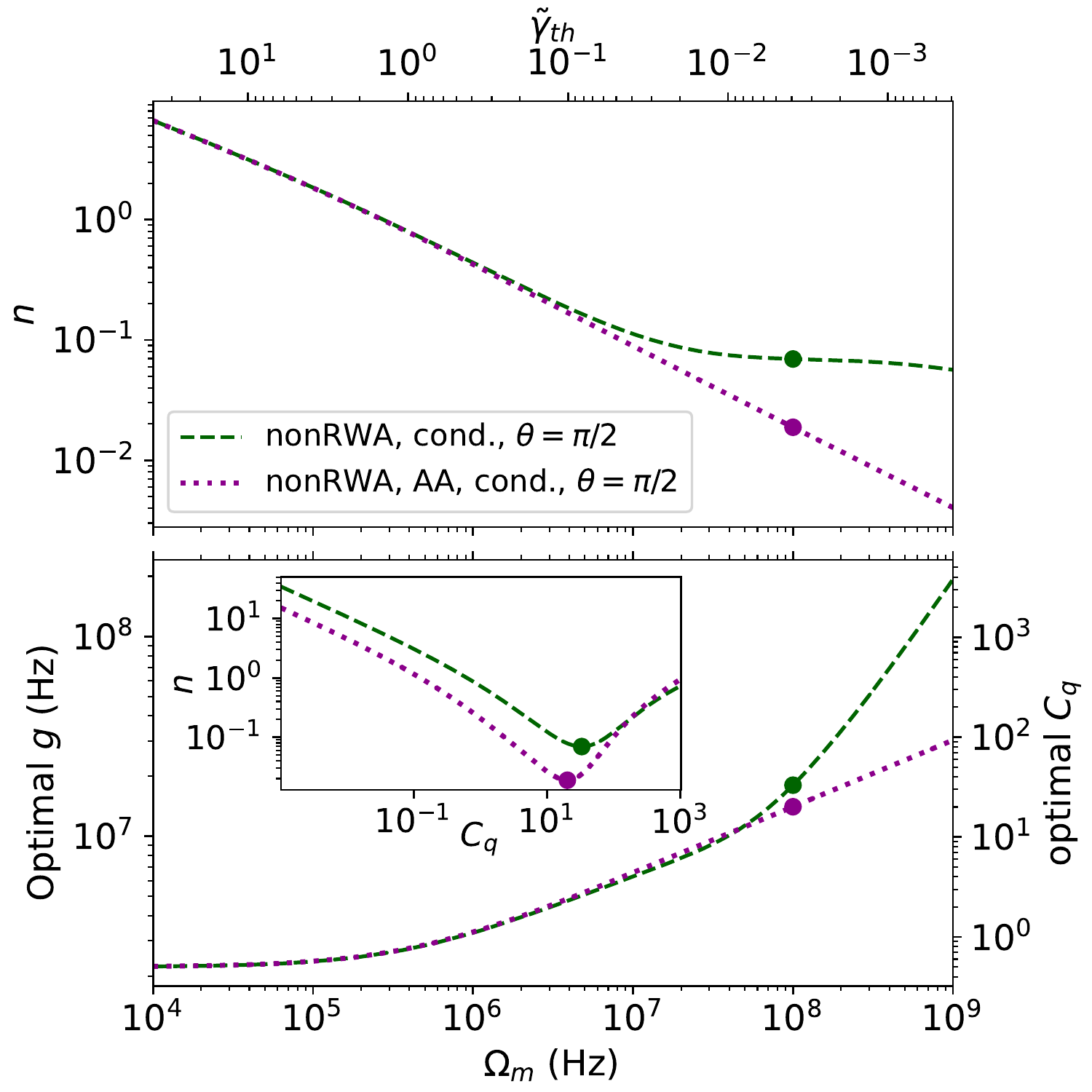}
	\caption{Plots of conditional (cond.) phonon number $n$ at the optimal coupling strength $g$ versus mechanical frequency $\Omega_m$ and normalized thermalization rate $\tilde{\gamma}_{th} \propto \Omega_m^{-1}$ for the nonRWA, with or without the adiabatic approximation (AA), for $\theta = \pi/2$. Top: Mechanical phonon number $n$ is plotted. Bottom: Plots of the optimal value of $C_q$. Inset: Conditional phonon number $n$ vs $C_q$ for $\Omega_m = 10^8$ Hz. circular dots on the plots mark points of optimal $C_q$ for $\Omega_m = 10^8$ Hz. Parameters used in this figure are $\kappa = \SI{e8}{Hz}$, $Q_m = 10^8$, and $\eta = 1$.}
	\label{fig:n-vs-Omega_m-adiabadic-comparison}
\end{figure}

We finally consider the effect that the adiabatic approximation might have on phonon occupancy. To illustrate this, in Fig.~\ref{fig:n-vs-Omega_m-adiabadic-comparison} we plot the conditional phonon occupancy as a function of the frequency and thermalization rate both for the exact solution and for the one applying the adiabatic approximation. In this plot, we have optimized the coupling strength but set the measurement angle to $\pi/2$. As expected, the adiabatic approximation breaks down when the mechanical frequency approaches the cavity bandwidth which is set to $\kappa=10^8$ Hz. The effect is further illustrated in the inset where we plot the phonon occupancy against the coupling strength.

\subsection{Mechanical squeezing}\label{seq:results-squeezing}

In this section we investigate the potential of generating a squeezed state of the mechanical oscillator, conditionally and unconditionally, without resorting to the conventional RWA and the adiabatic approximation. While the analysis of generating squeezed mechanical state without resorting to the RWA has already been performed by Meng et al \cite{Meng2020}, here we will extend the analysis by considering the effect of optimizing the measurement angle, $\theta$, exploring the validity of the adiabatic approximation and estimating the unconditional quantum state including feedback.

\begin{figure}[ht]
	\centering
	\includegraphics[width = \linewidth]{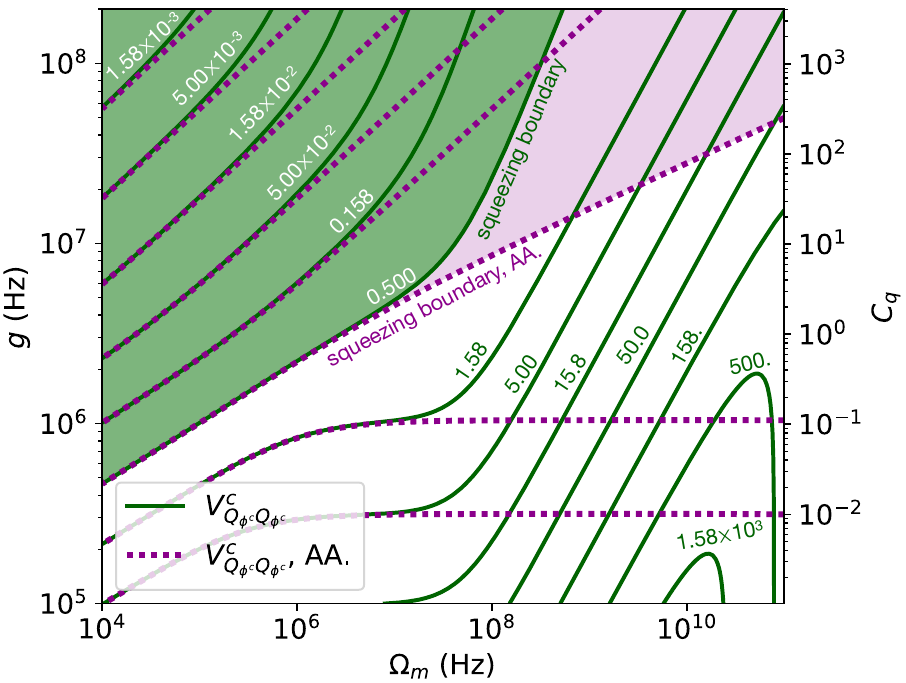}
	\caption{Contour plot of the conditional minimum variance $V^c_{Q_{\phi^c}Q_{\phi^c}}$ as a function of $\Omega_m$ and $g$, using the exact nonRWA model (green solid contour lines) or the nonRWA with the adiabatic approximation (AA.) from Ref.~\cite{Meng2020} (purple dotted lines). Marked on the plot are the values of $V^c_{Q_{\phi^c}Q_{\phi^c}}$ that each of the contours correspond to. The regions at which each of the two models predict mechanical squeezing ($V^c_{Q_{\phi^c}Q_{\phi^c}} < 0.5$) are coloured in with dark green or light pink, respectively. Parameters used in this plot are $\theta = \pi/2$, $\kappa = \SI{e8}{Hz}$, $Q_m = 10^8$, and $\eta = 1$.}\label{fig:Vc_Qphi-vs-Omega_m-and-g-MB2020-comparison}
\end{figure}

We start by illustrating the shortcomings of the adiabatic approximation in Fig.~\ref{fig:Vc_Qphi-vs-Omega_m-and-g-MB2020-comparison}. Here we plot the conditional minimum variance as a function of the coupling strength and the mechanical frequency with and without applying the adiabatic approximation. We observe that the two models completely agree for low frequencies and coupling strengths, but deviate when either $\Omega_m$ or $g$ approaches or exceeds $\kappa$. In these cases the adiabatic approximation overestimates the achievable minimum variance compared to the full solution. E.g. for $g=\SI{e7}{Hz}$ and $\Omega_m=\SI{e8}{Hz}$, the exact solution does not predict squeezing ($V^c_{Q_{\phi^c}Q_{\phi^c}} = 0.61$) while the approximate solution does ($V^c_{Q_{\phi^c}Q_{\phi^c}}=0.48$).    
At the breakdown of the adiabatic approximation, we observe a significant increase in the minimum variance for increasing $\Omega_m$. This is caused by the fact that the measured output field has interacted with the mechanics over multiple mechanical periods, thereby giving less timely information about the mechanical state. On the other hand, the minimum variance decreases again at even higher frequencies, where the number of thermal phonons is small.

\begin{figure}[ht]
	\centering
	\includegraphics[width = \linewidth]{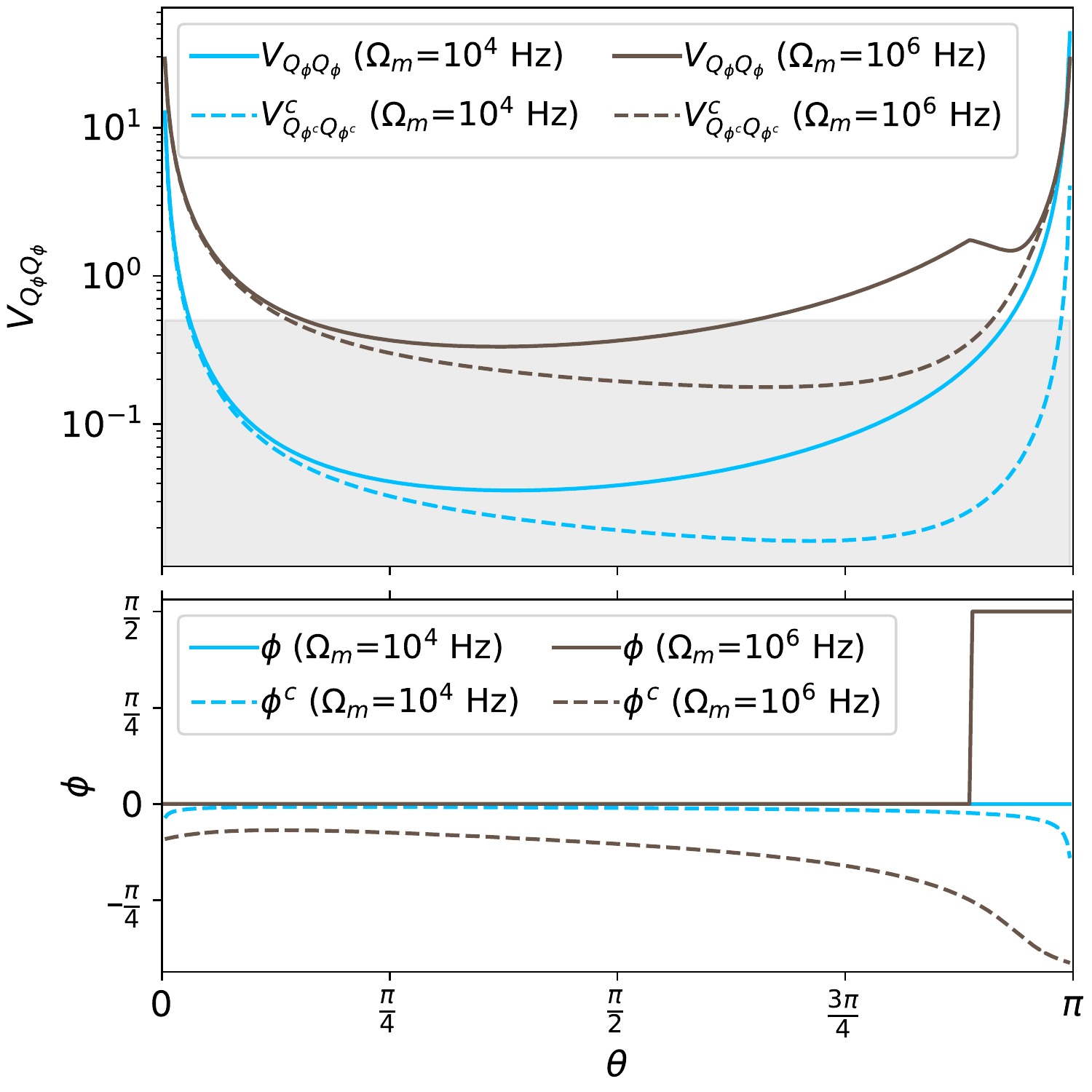}
	\caption{Top: Plot of the minimum variance of the mechanics vs measurement angle $\theta$ for both the conditional ($V^c_{Q_{\phi^c}Q_{\phi^c}}$) and the unconditional case ($V_{Q_{\phi}Q_{\phi}}$) for $\Omega_m = 10^4$ Hz and $\Omega_m = 10^6$ Hz. The grey-shaded area indicates mechanical squeezing. Bottom: squeezing angles vs $\theta$ for the conditional ($\phi^c$) and the unconditional case ($\phi$). Parameters used for these plots are $g = \SI{5e6}{Hz}$, $Q_m = 10^8$, $\kappa = 10^8$ Hz, and $\eta = 1$.}
	\label{fig:VQmin-vs-theta}
\end{figure}

Our next step is to analyze the effect of optimizing the measurement angle, $\theta$, for maximizing the degree of mechanical squeezing. In Fig.~\ref{fig:VQmin-vs-theta} top, the conditional and unconditional minimum variances $V^c_{Q_{\phi^c}Q_{\phi^c}}$ and $V_{Q_{\phi}Q_{\phi}}$, respectively, are plotted against $\theta \in [0,\pi]$ for $\Omega_m = \SI{e4}{Hz}$ and $\Omega_m = \SI{e6}{Hz}$. 
It is clear that squeezing can be produced for a rather large range of local oscillator phases (for this particular choice of parameters), and that the phase for which optimum squeezing is achieved is very different for the conditional and unconditional states. Interestingly, we also see that the squeezing angle $\phi$ varies continuously with the measurement angle for the conditional state while it stays constant at $\phi=0$ for the unconditional state. This means that no matter what light quadrature is being measured, the position variable of the mechanics will always attain the smallest variance among all mechanical quadratures for the unconditional state. It is however interesting to note that for other choices of the parameters (e.g. choosing a larger mechanical frequency), the smallest variance will occur in the momentum variable ($\phi = \pi/2$) for certain measurement angles. The changeover from $\phi = 0$ to $\phi = \pi/2$ happens when the conditional squeezing angle $\phi_c$ reaches $-\pi/4$, which never happens under the adiabatic approximation \cite{Meng2020}. This is shown in Fig.~\ref{fig:VQmin-vs-theta}, bottom (for $\Omega_m = 10^4$ Hz and $\Omega_m = 10^6$ Hz), where it is clear that the angle corresponding to the minimized variance is a binary function of the measurement angle. Furthermore, for this choice of the mechanical frequency, the range of measurements angles for which squeezing is observed is strongly reduced. The phase space nature of the unconditional state is thus very different from that of the conditional state.

\begin{figure}[ht]
\centering	
\includegraphics[width = \linewidth]{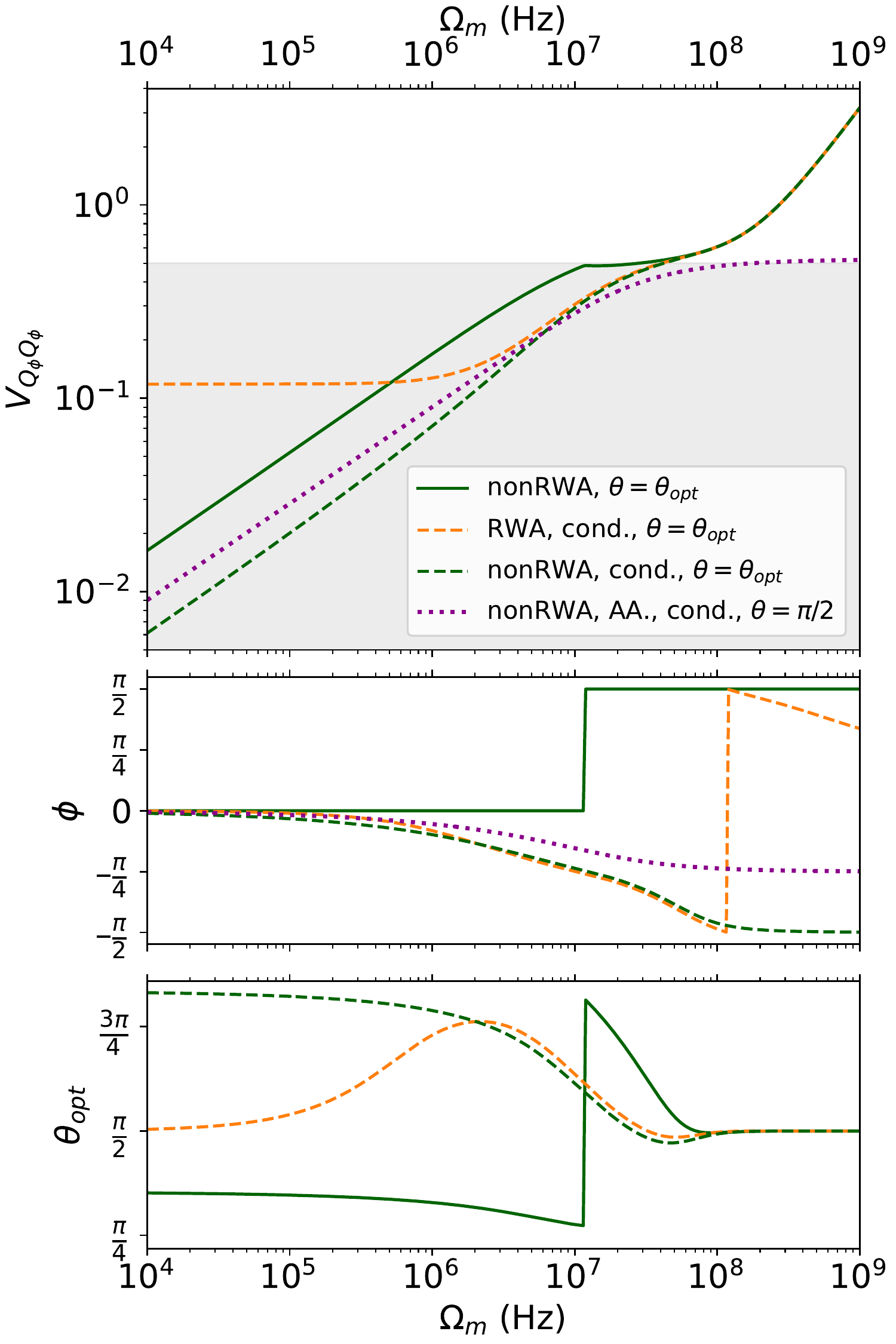}
\caption{Plot of the minimum mechanical variance   versus $\Omega_m$ in both the conditional (cond., dashed lines) and the unconditional case (solid lines) for the RWA (light orange) and nonRWA (dark green) model with measurement angles $\theta = \theta_{\text{opt}}$. Also included is the conditional nonRWA model with the adiabatic approximation (AA.) from Ref.~\cite{Meng2020} with $\theta = \pi/2$ (purple dotted lines). Top: Minimum mechanical variance is plotted. The grey-shaded area indicates mechanical squeezing. Middle: corresponding values of $\phi$ and $\phi^c$ is plotted. Bottom: corresponding values of $\theta_{\text{opt}}$ is plotted. Parameters used for these plots are $g = 10^7$ Hz, $Q_m = 10^8$ Hz, $\kappa = 10^8$ Hz, and $\eta = 1$.}\label{fig:Asymp-Vqmin-theta_opt-phi-vs-omega-g-is-1e7}
\end{figure}

In Fig.~\ref{fig:Asymp-Vqmin-theta_opt-phi-vs-omega-g-is-1e7} we plot the minimized mechanical variance as a function of the mechanical frequency for the conditional and unconditional states where the variance has been minimized over the measurement angle, $\theta$ (with the optimal angle called $\theta_{\text{opt}}$) while the quality factor and coupling strengths are kept constant. It is clear that 
for mechanical oscillators in the low frequency regime, a significant amount of squeezing can be observed for both the conditional and the unconditional case. These squeezing amounts are diminished if one applies the rotating wave approximation to the environment as illustrated by the light-orange dashed curve.  
In the high frequency regime (at the order of the cavity bandwidth or larger), the cavity dynamics will average out the dynamics of the mechanics, thereby smearing out the squeezing effect. In this case the variances of the conditional and unconditional states overlap, and moreover, the rotating wave approximation to the environment becomes valid.

Finally we make a comparison to the conditional solution found in Ref. \cite{Meng2020} where the adiabatic approximation was applied and the measurement angle was set to $\pi/2$. This solution is represented by the purple dotted curves in Fig.~\ref{fig:Asymp-Vqmin-theta_opt-phi-vs-omega-g-is-1e7}, and it is clear that the adiabatic approximation breaks down at large frequencies as expected, and that the degree of attainable conditional squeezing is underestimated.

\section{Experimental feasibility}\label{sec:feasibility}

Achieving ground state cooling at room temperature using this protocol appears within reach of currently existing technology. For example, using the parameters reported in \cite{Rossi2018} of $\Omega_m = 2\pi\cdot\SI{1.139}{MHz}$, $Q_m = 1.03\cdot 10^9$, $\kappa = 2\pi\cdot\SI{15,9}{Mhz}$, and $\eta = 0.77$, and adjusting the largest reported value of $C_q = 7.8$ at $T=\SI{4}{K}$ to the equivalent value at $T = \SI{300}{K}$ (corresponding to $g = \SI{3.1e5}{Hz}$) yields an unconditional phonon number (at $p/q \rightarrow \infty$) of $n = 1.38$ regardless of whether the RWA or nonRWA model is used. For a discussion on how much feedback strength is needed, see Appendix~\ref{app:finite-feedback}. Reaching the ground state $n < 1$ requires approximately $g > \SI{4e5}{Hz}$, which can be achieved by increasing the input laser power or the single-photon coupling strength of the oscillator. Observing the difference between the RWA and the nonRWA model will require a lower-frequency oscillator and/or a lower quality factor to get into the domain represented in \cref{fig:n-and-theta-opt-vs-g}. This might for example be done using the lower-frequency oscillators reported in Ref.~\cite{Hoej2021}. These oscillators might also be used to demonstrate mechanical squeezing as illustrated in \cref{fig:VQmin-vs-theta}, although this will largely depend on the achievable coupling strength.

All this being said, a number of experimental challenges may inhibit the realization of this control scheme. First, the model parameters must be determined to a high precision, as it is well known from classical control theory that imperfect knowledge of the model parameters may compromise the stability of the system. The desired precision will depend on the exact details on the experiment. It may also be necessary to account for extra noise sources, such as classical laser noise. Alternatively, it would be interesting to modify the model by employing techniques of robust control,  see e.g. Ref.~\cite{Yamamoto2006} where model uncertainty is built into the control scheme. Also, the model may need to be modified if there is an appreciable delay time between the measurement and feedback. In general, one needs the delay time to be at least significantly smaller than the mechanical period, which requires fast hardware for the estimation and control steps. If this delay time cannot be ignored, the state space model may need to be extended, see for instance Ref.~\cite[Ch.~6]{Wiseman2010}. We refer the reader to Refs.~\cite{Wieczorek2015,Magrini2021} in which experimental realizations of similar models have been done for further discussions on these issues.

\section{Conclusion}\label{sec:conclusion}
In conclusion, we have developed the formalism for optimal control of a mechanical oscillator without residing to any rotating-wave approximations for the mechanical oscillator. Using this formalism we discuss the resulting optomechanical dynamics with measurement and feedback, and deduced the minimal phonon occupancy as well as the minimal quadrature variance of the controlled mechanical oscillator. This was done both for the conditional state (where the measurement record is used to infer the state) and the unconditional state (where the measurement record is actively used to steer the mechanical oscillator into the desired state). We find that the rotating wave approximation of the mechanics to the environment is not valid in a rather large parameter space that is feasibly accessible in current optomechanical systems. Furthermore, we find that as a result of decoherence of the mechanical oscillator during feedback, the purity of the unconditional state is degraded compared to the conditional state. However, ground state cooling and squeezing of the unconditional state is still attainable in a room temperature environment. Indeed, ground state cooling of a room-temperature mechanical oscillator has recently been demonstrated \cite{Magrini2021}. As a final note, we remark that while we have worked with the most widely used non-rotating-wave approximation, a very interesting future study would be to apply some of the alternative models that guarantees positivity of the density matrix from e.g. Refs.~\cite{Diosi1993,Diosi1993b,Gao1997,Vacchini2000,Giovannetti2001,Barchielli2015}  in order to see if they predict different results. 


\appendix


\section{Linearisation of the optomechanical Hamiltonian including Feedback}\label{sec:linearisation}
To ensure consistency with the full nonlinear cavity-optomechanical  Hamiltonian including dissipation and measurement, we here derive from it the linearized Hamiltonian in Eq.~\eqref{eq:hamiltonian}. The section is inspired by many similar derivations in the literature, see e.g. Ref.~\cite{Hofer2011}.

The full Hamiltonian (disregarding the harmonic oscillator zero-point energies, and in a frame rotating with the laser frequency $\omega_L$ relative to the cavity field) is \cite{Law1995}
\begin{equation}
\begin{split}
	\fhat{H} = \hat{H}(t) =& \hbar\Delta_0\hat{a}^{\dagger}\hat{a} + \hbar\Omega_m\hat{b}^{\dagger}\hat{b} + \hbar g_0\hat{a}^{\dagger}\hat{a}(\hat{b} + \hat{b}^{\dagger}) 
	\\
	&+ i\hbar\sqrt{\kappa}\left[\drivefield(t)\hat{a}^{\dagger} - \drivefield^*(t)\hat{a} \right],
\end{split}
\end{equation}
where $\Delta_0 = \omega_c - \omega_L$ is the detuning of the input field, and $g_0$ is the vacuum optomechanical coupling rate.
In order to make the linearisation approximation, we perform the following time-dependent transformation of the density matrix: 
\begin{equation}
\dmatrix \rightarrow \bar{\rho} = U\dmatrix U^{\dagger},
\end{equation}
with
\begin{equation}
U = U(t) = D_a(-\alpha(t))D_b(-\beta(t))	
\end{equation}
where $ D_d(\delta) = e^{\delta \fhat{d}^{\dagger} - \delta^*\fhat{d}}$ is the standard displacement operator, defined for some annihilation operator $ \fhat{d} $ and complex number $ \delta $. This transforms the Schrödinger picture operators $\fhat{a}$ and $\fhat{b}$ to 
\begin{equation}
\begin{split}
    \bar{a}(t) &:= U(t)\fhat{a}U^{\dagger}(t) = \fhat{a} + \alpha(t) \\
    \bar{b}(t) &:= U(t)\fhat{b}U^{\dagger}(t) = \fhat{b} + \beta(t),
\end{split}
\end{equation}
which can be shown using the Baker-Hausdorff lemma. 
Note that we will frequently drop the time argument of the $\alpha$ and $\beta$ functions in the following.
The density matrix evolves according to the master equation:


\begin{equation}\label{eq:linearisation-meq}
\begin{split}
	d\dmatrix = &-\frac{i}{\hbar}[\Ham,\dmatrix]dt + \Lenv\dmatrix, dt \\
	&+ \kappa\mathcal{D}[\fhat{a}]\dmatrix dt + \sqrt{\eta\kappa}\mathcal{H}[\fhat{a}e^{-i\theta}]\dmatrix dW
\end{split}
\end{equation}
where the superoperators $\mathcal{D}$, $\mathcal{H}$ and $\Lenv$ are defined as in Sec.~\ref{sec:model}, i.e.  $\Lenv$ either equals $\LRWA$ or $\LnonRWA$ depending on which model is used.

We will now determine a new Hamiltonian $\bar{H}$ such that the master equation \eqref{eq:linearisation-meq} is fulfilled under the substitutions  $\dmatrix \rightarrow \bar{\rho}$ and $\fhat{H} \rightarrow \bar{H}$.

The master equation of $\bar{\rho}$ is as follows:

\begin{equation}
	\begin{split}
		d\bar{\rho} &= \bar{\rho}(t+dt)-\bar{\rho}(t) 
		\\
		&= U(t+dt)\dmatrix(t+dt) U^{\dagger}(t+dt) - U(t)\dmatrix(t) U^{\dagger}(t)  
		\\
		&= dU(t)\dmatrix(t)U^{\dagger}(t) + U(t)d\dmatrix U^{\dagger}(t) + U(t)\dmatrix dU^{\dagger}  
		\\
		&= Ud\dmatrix U^{\dagger} + [\dot{\alpha}^*\fhat{a}-\dot{\alpha}\fhat{a}^{\dagger} + \dot{\beta}^*\fhat{b} -  \dot{\beta}\fhat{b}^{\dagger},\bar{\rho}]dt,
	\end{split}
\end{equation}
where we have used that $\frac{d}{dt}D_a(-\alpha(t)) = [(\dot{\alpha}^*\fhat{a}-\dot{\alpha}\fhat{a}^{\dagger})+\frac{1}{2}(\dot{\alpha}^*\alpha-\dot{\alpha}\alpha^*)]D_a(-\alpha(t)$, and similarly for $D_b(-\beta(t))$. These derivatives are found e.g. by using the Taylor expansions of the displacement operator.

%
%


The transformation of each of the terms in the expression of $U d\dmatrix U^{\dagger}$ is:

\begin{align}
	U(\mathcal{D}[\fhat{a}])\dmatrix U^{\dagger} &= \mathcal{D}[\fhat{a}]\bar{\rho} - \frac{1}{2}[\alpha \fhat{a}^{\dagger}-\alpha^*\fhat{a},\bar{\rho}]
	\\
	U(\LRWA\dmatrix) U^{\dagger} &= \LRWA\bar{\rho} - \frac{\Gamma_m}{2}  [\beta \fhat{b}^{\dagger}-\beta^*\fhat{b},\bar{\rho}] 
	\\
	U(\LnonRWA\dmatrix) U^{\dagger} &= \LnonRWA\bar{\rho} -\frac{\Gamma_m}{2}[(\beta-\beta^*)(\fhat{b} +  \fhat{b}^{\dagger}),\bar{\rho}]
	\\
	U(\mathcal{H}[\fhat{a}e^{-i\theta}]\dmatrix)U^{\dagger} &= \mathcal{H}[\fhat{a}e^{-i\theta}]\bar{\rho}
 \end{align}

 Thus, putting everything together, we obtain:


\begin{equation}
\begin{split}
	\frac{d}{dt}\bar{\rho} = -&\frac{i}{\hbar}
	[U\fhat{H}U^{\dagger},\bar{\rho}] + \Lenv\bar{\rho} dt \\
	&+ \kappa\mathcal{D}[\fhat{a}]\bar{\rho} dt + \sqrt{\eta\kappa}\mathcal{H}[\fhat{a}e^{-i\theta}]\bar{\rho} dW \\
	+ &\left[\frac{\kappa}{2}(\alpha^*\fhat{a}-\alpha
	\fhat{a}^{\dagger}) + \mathcal{F}_{\text{env}} \right.
	\\
	 &\left. +  \dot{\alpha}^*\fhat{a}-\dot{\alpha}
	\fhat{a}^{\dagger} + \dot{\beta}^*\fhat{b} -  \dot{\beta}
	\fhat{b}^{\dagger},\bar{\rho}\right]dt
\end{split}
\end{equation}
with $\mathcal{F}_{\text{env}}$ equaling $\mathcal{F}_{\text{RWA}}$ or $\mathcal{F}_{\text{nonRWA}}$ defined below:
\begin{align}
	\mathcal{F}_{\text{RWA}} &=  \frac{\Gamma_m}{2}(\beta^*\fhat{b} - \beta \fhat{b}^{\dagger})
	\\
	\mathcal{F}_{\text{nonRWA}} &= \frac{\Gamma_m}{2}(\beta^*-\beta)(\fhat{b} +  \fhat{b}^{\dagger})
\end{align}

This yields an expression for $\bar{H}$:

\begin{equation}
\begin{split}
	\bar{H} = U\fhat{H}U^{\dagger} + i\hbar &\left\{ \frac{\kappa}{2}(\alpha^*\fhat{a}-\alpha \fhat{a}^{\dagger}) + \frac{\Gamma_m}{2}(\beta^*\fhat{b} - \beta \fhat{b}^{\dagger}) \right.
	\\
	 & \left.+ \dot{\alpha}^*\fhat{a}-\dot{\alpha}\fhat{a}^{\dagger} + \dot{\beta}^*\fhat{b} -  \dot{\beta}\fhat{b}^{\dagger}\right\}
\end{split}
\end{equation}

Evaluating $U\fhat{H}U^{\dagger}$ is straightforward, if a bit tedious. In the end we find that, when neglegting terms that are purely complex numbers (and therefore do not add to the dynamics), $\bar{H}$ can be written as
\begin{equation}\label{eq:linearisation-h-trans-done}
\begin{split}
	\bar{H} = &\hbar\Delta \fhat{a}^{\dagger}\fhat{a} + \hbar\Omega_m\fhat{b}^{\dagger}\fhat{b}+ \hbar g(\fhat{a}^{\dagger}\fhat{b} + \fhat{a}^{\dagger}\fhat{b}^{\dagger}) + \hbar g^*(\fhat{a}\fhat{b} + \fhat{a}\fhat{b}^{\dagger})
	\\
	 &+  i\hbar\sqrt{\kappa}(\drivefield_{\text{fb}}\fhat{a}^{\dagger}-\drivefield_{\text{fb}}^*\fhat{a}) 
+ \hbar g_0\fhat{a}^{\dagger}\fhat{a}(\fhat{b} + \fhat{b}^{\dagger}),
\end{split}
\end{equation}
with effective detuning $\Delta := \Delta_0 + g_0(\beta + \beta^*)$ and effective coupling $g := g_0\alpha$, all as long as $\alpha$ and $\beta$ satisfies the following coupled differential equations:

\begin{subequations}\label{eq:linearisation-alpha-beta-diffeqs}
\begin{align}
\dot{\alpha} &= -i[\Delta_0 + g_0(\beta + \beta^*)]\alpha - \frac{\kappa}{2}\alpha + \sqrt{\kappa}\drivefield_{\text{probe}} \label{eq:linearisation-alpha-diffeq}\\
\dot{\beta} &= -i\Omega_m\beta +\mathcal{G}_{\text{env}} -ig_0|\alpha|^2,
\end{align}	
\end{subequations}
where $\mathcal{G}_{\text{env}}$ equals $\mathcal{G}_{\text{RWA}} = -\frac{\Gamma_m}{2}\beta$ or $\mathcal{G}_{\text{nonRWA}} = -\frac{\Gamma_m}{2}(\beta-\beta^*)$, depending on the approximation. Throughout the calculations in the the main body of the paper, we work in this displaced picture where $\alpha$ and $\beta$ are chosen to satisfy these equations. Note that if $\drivefield$ is substituted with $\drivefield_{\text{probe}}$, these values of $\alpha(t),\beta(t)$  are equal to the expectation values $\braket{\fhat{a}(t)}, \braket{\fhat{b} (t)}$ in the undisplaced picture. 

The existence and nature of steady-state solutions to Eqs.~\eqref{eq:linearisation-alpha-beta-diffeqs} are discussed in e.g. \cite{BowenMilburnOptomechanics}. In the steady state, for constant $\drivefield_{\text{probe}}$, $\Delta$ and $g$ are approximately constant. In some cases, $g$ will be real, such as when $\Delta$ is set to 0 as in this paper. For sufficiently strong driving strengths $\drivefield_{\text{probe}}$ and not too strong feedback strengths $\drivefield_{\text{fb}}$, the nonlinear term $\hbar g_0\fhat{a}^{\dagger}\fhat{a}(\fhat{b} + \fhat{b}^{\dagger})$ may be neglected from the Hamiltonian. Thus, under these conditions, we obtain the Hamiltonian in Eq.~\eqref{eq:hamiltonian}. The system may exert nonlinear instabilities for very high values of $g$, but the exact instability points will depend on the single-photon coupling strength $g_0$ \cite{Ludwig2008}. It should be noted that we have not given precise requirements or bounds on the values of $\drivefield_{\text{probe}}$ and  $\drivefield_{\text{fb}}$. These considerations are left for future studies.
\section{Deriving the equations of motion}\label{app:equations-of-motion}

When deriving the covariance matrix elements in Eqs.~\eqref{eq:conditional-state-space}, several smaller challenges must be dealt with:
First of all, denoting the covariance of two arbitrary operators $\fhat{A}$ and $\fhat{B}$ by $\text{Cov}[\fhat{A},\fhat{B}]$= $V_{AB} = \Real(\braket{ \fhat{A}\fhat{B}} )- \braket{\fhat{A}} \braket{\fhat{B}})$, we have that 
\begin{equation}\label{supp:dvab}
	dV_{AB} = \frac{1}{2}(d\langle \fhat{A} \fhat{B} \rangle + d\langle \fhat{B} \fhat{A} \rangle) -d(\langle \fhat{A} \rangle \langle \fhat{B} \rangle).
\end{equation}
The last term must be evaluated using the Itō Calculus product rule stating that for two functions $f$ and $g$ evolving according to the stochastic differential equations $df = f_1dt + f_2dW$ and $dg = g_1dt + g_2dW$, then:

\begin{equation}
	d(fg) = fdg + gdf + dfdg
\end{equation}
While in normal calculus the term $dfdg$ vanishes, as it is proportional to $dt^2$, here we have that
\begin{equation}
	\begin{split}
dfdg &= f_1g_1dt^2 + (f_1g_2 + g_2f_1)dtdW + f_2g_2dW^2 \\
&= f_2g_2dt
\end{split}
\end{equation}
since $dW^2 = dt$.

Secondly, the differential equations for the variances will contain third order moments of $\fhat{A}$ and $\fhat{B}$, i.e. $\langle \fhat{A}^2 \fhat{B} \rangle$. However, because the quantum state is assumed to be Gaussian, such products can be expressed in terms of first and second order moments. This property has been noted before in e.g. Refs.~\cite{Jacobs2006,BowenMilburnOptomechanics}. It can be shown by the use of \textit{Isserlis' theorem}. A specific instance of this theorem states that for three (classical) stochastic variables $X_1,X_2,X_3$ that follows a multivariate Gaussian distribution, it holds that \cite{Isserlis1918}
\begin{equation}\label{eq:isserlis-rearranged}
\begin{split}
	E[X_1X_2X_3] =& E[X_1X_2]\mu_3 + E[X_1X_3]\mu_2 + E[X_2X_3]\mu_1 \\&-2\mu_1\mu_2\mu_3
\end{split}
\end{equation}
where $E$ is the classical expectation value (strictly different from the expectation value $\mathrm{E}$ (non-italicized) used in the main text) defined by
\begin{equation}
	E[f(\mathbf{X})] = \int_{\mathbb{R}^n}f(\mathbf{x})p(\mathbf{x})d^n\mathbf{x},
\end{equation}
and $\mu_i = E[X_i],\: i = 1,2,3$ is the means of $X_i$. In the above formula, $p(\mathbf{x})$ is the probability density function corresponding to the stochastic vector $\mathbf{X}$ of length $n$. For a multivariate Gaussian distribution, 

\begin{equation}
	p(\mathbf{x}) = \frac{1}{\sqrt{(2\pi)^n\det{\mathbf{V}}}} \exp(-\frac{1}{2}(\mathbf{x}-\boldsymbol{\mu})^T\mathbf{V}^{-1}(\mathbf{x}-\boldsymbol{\mu}))
\end{equation}
where $\boldsymbol{\mu} = E[\mathbf{X}]$ and $\mathbf{V} = E[(\mathbf{X}-\boldsymbol{\mu})(\mathbf{X}-\boldsymbol{\mu})^T]$. When working with Gaussian states in quantum information, the above formula exactly equals the form of the Wigner function $W(\mathbf{x})$ of the state $\dmatrix$ (See \cite{Weedbrook2012} for proper definitions). For the Wigner function, however, the 'expectation value'  $E[...]$ is \textit{not} equal to the quantum expectation $\langle ... \rangle$. Rather, we have that for any function $f$, \cite{Leonhardt1998}

\begin{equation}
	E[f(\mathbf{X})] = \int_{\mathbb{R}^n}f(\mathbf{x})W(\mathbf{x})d^n\mathbf{x} = \langle \hat{F}(\hat{\mathbf{X}} \rangle)
\end{equation}
where, $\hat{F}$ is an operator function of the operators $\hat{\mathbf{X}}$, that symmetrizes $f$. Specifically, $\hat{F}$ satisfies the properties

\begin{enumerate}
	\item $\hat{F}(\mathbf{x}) = f(\mathbf{x})$
	\item $\hat{F}(P\hat{\mathbf{X}}) = \hat{F}(\hat{\mathbf{X}})$ For all permutations P.
\end{enumerate}
In the above, $P\hat{\mathbf{X}}$ represents a permutation of the elements in the vector of operators $\hatvek{X}$, e.g. swapping two elements.

As an example, if $f(\mathbf{x}) = x^2y$, then $\hat{F}(\hat{\mathbf{X}}) = \frac{1}{3}(\fhat{X}^2\fhat{Y} + \fhat{X}\fhat{Y}\fhat{X} + \fhat{Y}\fhat{X}^2)$, i.e. an average over all the ways the operators $(\fhat{X},\fhat{X},\fhat{Y})$ can be ordered in a non-commutative product. This gives us:

\begin{equation}
	\frac{1}{3}(\langle \fhat{X}^2\fhat{Y} \rangle + \langle \fhat{X}\fhat{Y}\fhat{X} \rangle + \langle \fhat{Y}\fhat{X}^2 \rangle) = E[X^2Y] 
\end{equation}
The above sum of third order moments of $\fhat{X}$ and $\fhat{Y}$ can now be expressed as a sum of products of first and second order moments by application of Eq.~\ref{eq:isserlis-rearranged}. Applying identities such as the one above when evaluating $dV_{AB}$ for $\fhat{A},\fhat{B} \in \{\fhat{Q},\fhat{P},\fhat{X},\fhat{Y}\}$ leads us to the equations of motion in Eq.~\eqref{eq:vc-matrix}.

\section{Positivity of the density matrix}\label{sec:app-positiviy}

A real, symmetric, positive-definite matrix $\vek{V}$ corresponds to the covariance matrix of a quantum state density matrix $\dmatrix$ if and only if \cite{Simon1994,Weedbrook2012}
\begin{equation}\label{eq:positivity}
	\vek{V} + \frac{i}{2}\vek{\Omega} \ge 0
\end{equation}
where $\vek{\Omega}$ is the symplectic form defined in Ref.~\cite[Eq.~(2)]{Weedbrook2012}. Thus, if a given covariance matrix $\vek{V}$ satisfies the above inequality, its associated density matrix must be positive, as this is one of the defining properties of a quantum density matrix. 

As mentioned in Sec.~\ref{sec:model}, the nonRWA model of the mechanical interaction with its environment as given in Eq.~\eqref{eq:mech-loss-nonRWA} can occasionally produce nonpositive density matrices, which will of course be an invalid result. Therefore, for all of the figures in the main text, we have checked that all of the underlying (numerically calculated) covariance matrices were verified to fulfil Eq.~\eqref{eq:positivity}.


\section{Asymptotic feedback}\label{app:asymptotic}

In this section we derive the excess covariance matrix elements for $p/q \rightarrow \infty$ presented in Sec.~\ref{sec:asymptotic-feedback}. We focus on the nonRWA model with feedback that minimize the phonon number. The derivations for the RWA model and for feedback that optimize squeezing are similar and therefore omitted. The following derivations are valid under the conditions $ \Gamma_m > 0, \kappa > 0, g > 0, \theta \in \mathbb{R}, p > 0, q> 0$, as well as the condition that the mechanical oscillator is underdamped, i.e. $\Gamma_m < 2 \Omega_m$.

We are given the algebraic Riccati equation for the infinite-horizon
optimal control problem:

\begin{equation}\label{eq:feedback-algebraic-riccati}
 \mathbf{P} + \mathbf{A}^T\mathbf{Y} + \mathbf{YA} - \mathbf{YBQ}^{-1}\mathbf{B}^T\mathbf{Y} = 0,
\end{equation}
where $\mathbf{A}$, $\mathbf{B}$, $\mathbf{P}$, $\mathbf{Q}$, and $\mathbf{Y}$ are defined as in the main text. We normalize the equation by writing the rates of the system in units of $\Omega_m$, i.e. by dividing through by $\Omega_m$ and then performing the substitution  $x \rightarrow x \Omega_m$ for $x \in \{\Omega_m,\Gamma_m,g,\kappa\}$. 

We solve for the $4\times 4$ matrix $\mathbf{Y}$ by finding eigenvalues and eigenvectors to the following $8\times 8$ Hamiltonian matrix $\mathbf{H}$ (not related to the energy operator):
\begin{equation}
\begin{split}
	\mathbf{H} &= \begin{bmatrix}
		\mathbf{A} & -\mathbf{B}\mathbf{Q}^{-1}\mathbf{B}^T \\
		-\mathbf{P} & -\mathbf{A}^T 
	\end{bmatrix} \\
	&= \begin{bmatrix}
		 0 & 1 & 0 & 0 & 0 & 0 & 0 & 0 \\
 -1 & -\Gamma_m & -2 g & 0 & 0 & 0 & 0 & 0 \\
 0 & 0 & -\frac{\kappa}{2} & 0 & 0 & 0 & -\frac{\kappa}{q} & 0 \\
 -2 g & 0 & 0 & -\frac{\kappa}{2} & 0 & 0 & 0 & -\frac{\kappa}{q} \\
 -p & 0 & 0 & 0 & 0 & 1 & 0 & 2 g \\
 0 & -p & 0 & 0 & -1 & \Gamma_m & 0 & 0 \\
 0 & 0 & 0 & 0 & 0 & 2 g & \frac{\kappa}{2} & 0 \\
 0 & 0 & 0 & 0 & 0 & 0 & 0 & \frac{\kappa}{2} \\
	\end{bmatrix}
\end{split}
\end{equation}

We find the following characteristic polynomial $ch(\lambda) = \det(\vek{H}-\lambda \vek{I})$:
\begin{equation}
\begin{split}
		ch(\lambda) = -\frac{1}{8q} \left(\frac{\kappa }{2}-\lambda \right) \left(\frac{\kappa }{2}+\lambda \right) \Big[-8 g \left(4 g
   \kappa  p-4 g \kappa  \lambda ^2 p\right)&
   \\
   -q \left(-\Gamma_m^2 \lambda
   ^2+\lambda ^4+2 \lambda ^2+1\right) (\kappa^2-(2\lambda)^2) \Big]&
\end{split}
\end{equation}

it is immediately apparent that 
\begin{equation}
\lambda_1^{\pm} = \pm \frac{\kappa}{2}
\end{equation}
are two roots to $ch(\lambda)$ and thus eigenvalues to $\vek{H}$. to find the rest of the eigenvalues, we divide out the roots and simplify the characteristic equation $ch(\lambda) = 0$ to find:
\begin{equation}\label{eq:karpol-red-sigma1}
\begin{split}
	&\left(ch(\lambda) = 0 \;\wedge\; \lambda \neq \lambda_1^{\pm} \right) 
	\\
	\iff &4g^2\kappa (\lambda^2-1)\\
	 &+ \frac{q}{p}[\lambda^4+\lambda^2(2-\Gamma_m^2)+1]\left[\lambda^2-\left(\frac{\kappa}{2}\right)^2\right]= 0
\end{split}
\end{equation}

Note that the above is a cubic equation in the variable $\sigma = \lambda^2$, so we can write it as 
\begin{equation}\label{eq:asymp-feedback-sigma-eq}
	4g^2\kappa (\sigma-1) + \frac{q}{p}(\sigma^2+\sigma(2-\Gamma_m^2)+1)\left[\sigma-\left(\frac{\kappa}{2}\right)^2\right]= 0.
\end{equation}

In principle, this equation can be solved using the standard formula for a cubic equation. We can however simplify the problem slightly, as we are only interested in the limiting case $q/p \rightarrow 0$. Now, the second term on the left hand side of the above equation is directly proportional to $q/p$ and therefore goes to zero as $q/p$ goes to zero. This suggests that the cubic equation has a root $\sigma = \sigma_2$ for which  $\sigma_2 \rightarrow 1$ as $q/p \rightarrow 0$. To prove this statement, write the root as $\sigma_2 := 1 + r_2$, where $r_2$, which we call the \textit{remainder} of $\sigma_2$, is a complex number depending on the coefficients of the equation in \eqref{eq:asymp-feedback-sigma-eq}, in particular on $q/p$. if we substitute $\sigma = 1 + r_2$ into \eqref{eq:asymp-feedback-sigma-eq}, we get:

\begin{equation}\label{eq:asymp-karpol-sigma2}
\begin{split}
	\Bigg\{4g^2\kappa r_2+ &\frac{q}{p}\Big[(1+r_2)^2+(1+r_2)(2-\Gamma_m^2)+1\Big]\\
	&\times\left[(1+r_2)-\left(\frac{\kappa}{2}\right)^2\right]\Bigg\}= 0
	\end{split}
\end{equation}

It is clear that if we set $q/p = 0$ we have that $r_2 = 0$. Note that since the roots of a polynomial depend continuously on its coefficients, $r_2$ is a continuous function of $q/p$. Moreover, continuously increasing $q/p$ from zero will also continuously change $r_2$ (possibly branching into more than one root). equivalently, there must exist at least one root for which its remainder $r_2$ goes to zero, as desired. We can also show how quickly $r_2$ goes to zero: multiplying by $p/q$ on both sides of Eq.~\eqref{eq:asymp-karpol-sigma2}, we have

\begin{equation}
\begin{split}
	\Bigg\{4g^2\kappa \left(\frac{p}{q}\right)r_2+ &\left[(1+r_2)^2+(1+r_2)(2-\Gamma_m^2)+1\right] \\
	&\times\left[(1+r_2)-\left(\frac{\kappa}{2}\right)^2\right]\Bigg\}= 0,
\end{split}
\end{equation}
which can be rearranged to show that
\begin{equation}
		\lim_{q/p \rightarrow 0} \left(\frac{p}{q}\right)r_2 = - \frac{(4-\Gamma_m^2)\left[1-\left(\frac{\kappa}{2}\right)^2\right]}{4g^2\kappa}.
\end{equation}

We reduce the characteristic equation further by dividing Eq.~\eqref{eq:karpol-red-sigma1} by $\sigma - \sigma_2$, which gives:

\begin{equation}
\begin{split}
	&\frac{1}{4} \Gamma_m^2 \left(\kappa ^2-4\right)+\frac{4 g^2 \kappa
    p}{q}-\frac{3 \kappa ^2}{4}\\
    &+r_2^2
     -\frac{1}{4} r_2
   \left(4 \Gamma_m^2+\kappa ^2-16\right) + 4
     \\
     &+\sigma  \left(-\Gamma_m^2-\frac{\kappa ^2}{4}+r_2+3\right)+\sigma ^2 = 0
\end{split}
\end{equation}

We find the roots of the above quadratic using the quadratic formula: The discriminant $d$ is
\begin{equation}
\begin{split}
	d = -16 g^2 \kappa \frac{p}{q}+\Gamma_m^4+\frac{\kappa ^4}{16}-&3
   r_2^2+\frac{1}{2} \kappa ^2 \left(-\Gamma_m^2+r_2+3\right)
   \\
    &+2 \Gamma_m^2 (r_2-1)-10
   r_2-7r_2.
\end{split}
\end{equation}

In the expression for $d$, the term $-16 g^2 \kappa p/q$ is the only term proportional to $p/q$, while the other terms are constant or proportional to $r_2$ or $r_2^2$. Therefore, for sufficiently large values of $p/q$, $d$ is negative, or at least has negative real part as we do not yet know whether or not $r_2$ is real. Thus, the remaining roots to the characteristic equation are
\begin{align}
\sigma_3 &= \frac{1}{2}\left(\Gamma_m^2+\frac{\kappa^2}{4}-r_2-3 - i \sqrt{-d}\right), 
\\
\sigma_4 &= \frac{1}{2}\left(\Gamma_m^2+\frac{\kappa ^2}{4}-r_2-3 + i \sqrt{-d}\right).
\end{align}

We can now deduce that for sufficiently large values of $p/q$, the re $r_2$ is real. This is because, when $r_2$ is sufficiently small, $\sigma_3$ and $\sigma_4$ are necessarily complex. Since the reduced characteristic equation \eqref{eq:asymp-feedback-sigma-eq} has real coefficients, any complex roots must appear in conjugate pairs. The remaining root $\sigma_2 = 1+ r_2$ therefore has to be real, which proves that $d$ is real and negative.

We now have that the eigenvalues to $\mathbf{H}$ with negative real part is $ \lambda_1 = -\frac{\kappa}{2}$, $\lambda_i = -\sqrt{\sigma_i}, i = 2,3,4$. Corresponding eigenvectors are denoted $\vek{v}_i$, $i = 1,2,3,4$. $\lambda_3$ and $\lambda_4$ are given by:

\begin{align}
	\lambda_3 &= -\sqrt{\sigma_3} = \sqrt{g} \sqrt[4]{\frac{\kappa  p}{q}}(-1+i)  + r_3
	\\
	\lambda_4 &= -\sqrt{\sigma_4} = \lambda_3^*
\end{align}
where $r_3$ is a remainder that goes to zero as $q/p$ goes to zero.

It is easy to show that an eigenvector corresponding to $\lambda_1$ is $\mathbf{v}_1 = [0\; 0\; 0\; 1\; 0\; 0\; 0 \; 0]^T$. Through Gauss-Jordan elimination of $\mathbf{H}-\lambda_i\mathbf{I}$, with $\vek{I}$ being the ($8 \times 8$) identity matrix, we find that
\begin{equation}
	\mathbf{v}_2 = 
\begin{bmatrix}
 -\frac{1}{p \sqrt{r_2+1}} \\
 \frac{1}{p} \\
 \frac{\frac{r_2+2}{\sqrt{r_2+1}}-\Gamma_m}{2 g p} \\
 -\frac{4 g}{-\kappa  p \sqrt{r_2+1}+2 p r_2+2 p} \\
 -\frac{\frac{\Gamma_m}{\sqrt{r_2+1}}+2}{\Gamma_m \sqrt{r_2+1}+r_2+2} \\
 \frac{r_2}{\sqrt{r_2+1} \left(\Gamma_m \sqrt{r_2+1}+r_2+2\right)} \\
 \frac{q \left(\frac{r_2+2}{\sqrt{r_2+1}}-\Gamma_m\right) \left(\sqrt{r_2+1}-\frac{\kappa
   }{2}\right)}{2 g \kappa  p} \\
 0 \\
\end{bmatrix} 
\end{equation}
\\

\begin{widetext}
\begin{align}
\vek{v}_3 & = 
\begin{bmatrix}
\frac{2 g}{-(1-i) \Gamma_m \sqrt{g} \sqrt[4]{\frac{\kappa 
   p}{q}}-2 i g \sqrt{\frac{\kappa  p}{q}}+r_3
   \left(\Gamma_m-(2-2 i) \sqrt{g} \sqrt[4]{\frac{\kappa 
   p}{q}}\right)+r_3^2+1} \\
 \frac{2 g}{\Gamma_m-(1-i) \sqrt{g} \sqrt[4]{\frac{\kappa 
   p}{q}}+\frac{1}{r_3-(1-i) \sqrt{g} \sqrt[4]{\frac{\kappa 
   p}{q}}}+r_3} \\
 -1 \\
 -\frac{8 g^2}{\left(-(2-2 i) \sqrt{g} \sqrt[4]{\frac{\kappa  p}{q}}+\kappa
   +2 r_3\right) \left(-(1-i) \Gamma_m \sqrt{g}
   \sqrt[4]{\frac{\kappa  p}{q}}-2 i g \sqrt{\frac{\kappa 
   p}{q}}+r_3 \left(\Gamma_m-(2-2 i) \sqrt{g}
   \sqrt[4]{\frac{\kappa  p}{q}}\right)+r_3^2+1\right)} \\
 -\frac{2 g p \left(-\Gamma_m-(2-2 i) \sqrt{g}
   \sqrt[4]{\frac{\kappa  p}{q}}+2 r_3\right)}{\left(-(1-i)
   \Gamma_m \sqrt{g} \sqrt[4]{\frac{\kappa  p}{q}}-2 i g
   \sqrt{\frac{\kappa  p}{q}}+r_3 \left(\Gamma_m-(2-2 i)
   \sqrt{g} \sqrt[4]{\frac{\kappa  p}{q}}\right)+r_3^2+1\right)
   \left((1-i) \Gamma_m \sqrt{g} \sqrt[4]{\frac{\kappa  p}{q}}-2 i
   g \sqrt{\frac{\kappa  p}{q}}-r_3 \left(\Gamma_m+(2-2
   i) \sqrt{g} \sqrt[4]{\frac{\kappa 
   p}{q}}\right)+r_3^2+1\right)} \\
 -\frac{2 g p \left(-2 i g \sqrt{\frac{\kappa  p}{q}}-(2-2 i) \sqrt{g}
   r_3 \sqrt[4]{\frac{\kappa 
   p}{q}}+r_3^2-1\right)}{\left(-(1-i) \Gamma_m \sqrt{g}
   \sqrt[4]{\frac{\kappa  p}{q}}-2 i g \sqrt{\frac{\kappa 
   p}{q}}+r_3 \left(\Gamma_m-(2-2 i) \sqrt{g}
   \sqrt[4]{\frac{\kappa  p}{q}}\right)+r_3^2+1\right)
   \left((1-i) \Gamma_m \sqrt{g} \sqrt[4]{\frac{\kappa  p}{q}}-2 i
   g \sqrt{\frac{\kappa  p}{q}}-r_3 \left(\Gamma_m+(2-2
   i) \sqrt{g} \sqrt[4]{\frac{\kappa 
   p}{q}}\right)+r_3^2+1\right)} \\
 \frac{q \left(-(2-2 i) \sqrt{g} \sqrt[4]{\frac{\kappa  p}{q}}+\kappa +2
   r_3\right)}{2 \kappa } \\
 0 \\
 \end{bmatrix}
 \\
 \vek{v}_4 & = \vek{v}_3^*
\end{align}

\end{widetext}
These expressions all hold for sufficiently small values of $r_2$ and $r_3$. 
%
%

The algebraic Ricatti equation \eqref{eq:feedback-algebraic-riccati} is now solved as follows: Define an $8 \times 4$ matrix $\vek{U}$ and two $4 \times 4$ matrices $\vek{U}_1$ and $\vek{U}_2$ by

\begin{equation}
	\vek{U} :=: \begin{bmatrix}
		\vek{U}_1 \\ \vek{U}_2
	\end{bmatrix} := \begin{bmatrix}
		\vek{v}_1 & \vek{v}_2 & \vek{v}_3 & \vek{v}_4 \end{bmatrix}.
\end{equation}
Then it holds that \cite{Laub1978}
\begin{equation}
	\vek{Y} = \vek{U}_2\vek{U}_1^{-1}
\end{equation}
is the unique stabilizing solution to \eqref{eq:feedback-algebraic-riccati}, i.e. the unique solution that renders the matrix $\vek{A} - \vek{B}\vek{Q}^{-1}\vek{B}^T\vek{Y}$ asymptotically stable (eigenvalues with all negative real parts) \cite{Wiseman2010}.

We find that $\vek{Y}$ is of the form 

\begin{equation}
	\vek{Y} = \begin{bmatrix}
		y_{11} & y_{12} & y_{13} & 0 \\
		y_{12} & y_{22} & y_{23} & 0 \\
		y_{13} & y_{23} & y_{33} & 0 \\
		0 & 0 & 0 & 0
	\end{bmatrix}
\end{equation}

The above entries $y_{ij}$ are all complicated expressions of the system parameters and $r_2$ and $r_3$, and will not be written in full here. We are only interested in their behavior as $p/q$ goes to infinity. Note that we only care about the terms $y_{13}$, $y_{23}$, and $y_{33}$, as they are the only terms appearing in the Kalman Gain matrix $\vek{K} = \vek{Q}^{-1}\vek{B}^T\vek{Y} = \frac{\sqrt{\kappa}}{q}\begin{bmatrix}
	y_{13} & y_{23} & y_{33} & 0 \\
	0 & 0 & 0 & 0
\end{bmatrix}$. We find that


\begin{align}
	y_{13} &= -q\left(\sqrt{\frac{p}{\kappa q}} + O\left(\sqrt[4]{\frac{p}{q}}\right)\right)\\
	y_{23} &= -q\left(\sqrt{\frac{p}{\kappa q}} + O\left(\sqrt[4]{\frac{p}{q}}\right)\right) \\
	y_{33} &= 2q\left(\sqrt{\frac{g}{\kappa}}\sqrt[4]{\frac{p}{\kappa q}} + O(1)\right) 
\end{align}
The lower-order terms of $\vek{K}$ are not important; note however that, from the fact that $\vek{Y}$ is the unique stabilizing solution, it follows that $\vek{Y}$ is real, and thus also $\vek{K}$ is real. 

With the above expression for $\vek{K}$, we can now solve the Lyapunov equation for the excess covariance matrix $\V{\vek{X}}^E$:

\begin{equation}\label{eq:asymp-feedback-lyap}
	\vek{N}\V{\vek{X}}^E+\V{\vek{X}}^E\vek{N}^T + \vek{F}^T\vek{F} = 0
\end{equation}
with $\mathbf{N} = \mathbf{A} - \mathbf{BQ}^{-1}\mathbf{B}^T\mathbf{Y} $ and $\mathbf{F} = \mathbf{C}\Vc{\vek{X}}+\mathbf{\Gamma}$.
Equation \eqref{eq:asymp-feedback-lyap} is a linear system of equations in the entries of $\vek{V}_E$ and are therefore straightforwardly solved. We then take the limit of the solutions to these equations as $p/q \rightarrow \infty$ to obtain the expressions in Sec.~\ref{sec:asymptotic-feedback}.


\section{Finite feedback strength}\label{app:finite-feedback}
In this section we investigate the unconditional state as a function of the feedback strength. In the main text, we are primarily concerned with the state in the limit $p/q \rightarrow \infty$. One may however ask how large a value of $p/q$ is necessary in order that the unconditional state is 'close enough' to its limit value. This is relevant since if the feedback control field $\epsilon_{\text{fb}}$ is comparable to or exceeds the probe strength $\epsilon_{\text{probe}}$, the linearisation in Appendix~\ref{sec:linearisation} is questionable and the results shown in the main text are likely not experimentally realisable.

Under the optimal control scheme, the feedback is set to $\vek{u} = -\vek{K}\braket{\hatvek{X}}_c$. Its mean value is $\E{\vek{u}} = -\vek{K}\E{\braket{\hatvek{X}}_c} = \vek{0}$, while its variance is given by

\begin{equation}
	\V{\vek{u}} = \E{\vek{u}(\vek{u})^T} = \vek{K}\V{\vek{X}}^E\vek{K}^T.
\end{equation}
Since $y_{\text{\text{fb}}} = 0$, the only non-zero entry of the covariance matrix $\V{\vek{u}}$ is $V_{x_{\text{fb}}x_{\text{fb}}}$, the first diagonal entry. We define the \textit{feedback strength} as $\sigma_{\epsilon_{\text{fb}}} = \sqrt{V_{x_{\text{fb}}x_{\text{fb}}}}$, i.e. the standard deviation of $x_{\text{fb}}$. Note that we use  the symbol $\sigma_{\epsilon_{\text{fb}}}$ instead of $\sigma_{x_{\text{fb}}}$ as indeed,  $\epsilon_{\text{fb}} = x_{\text{fb}}$ (since $\epsilon_{\text{fb}} =  x_{\text{fb}} + iy_{\text{fb}}$ and $y_{\text{fb}} = 0$). Note that $\sigma_{\epsilon_{\text{fb}}} \rightarrow \infty$ for $p/q \rightarrow \infty$. This is because $\vek{K}$ increases as $\sqrt{p/q}$ to leading order (see e.g. Eq.~\eqref{eq:asymp-kalman-gain}) and $\V{\vek{X}}^E$ does not go to zero as $p/q \rightarrow \infty$.

To demonstrate how large the feedback strength needs to be in order that the limiting unconditional state is approximately realized, we use as an example the parameters from Ref.~\cite{Rossi2018} used in Sec.~\ref{sec:feasibility}.

From \cref{eq:linearisation-alpha-diffeq}, one finds in the steady state with the effective detuning $\Delta = 0$ that
\begin{equation}
	g = g_0\frac{2\epsilon_{\text{probe}}}{\sqrt{\kappa}} \iff \epsilon_{\text{probe}} = \frac{g}{g_0}\frac{\sqrt{\kappa}}{2}.
\end{equation}
Using the reported value in Ref.~\cite{Rossi2018} of the single  photon coupling strength $g_0 = 2\pi\cdot\SI{127}{Hz}$ and the coupling strength $g=\SI{3.1e5}{Hz}$ yields $\epsilon_{\text{probe}} = \SI{2.0e6}{Hz}$. 

\begin{figure}[ht]
    \centering
    \includegraphics[width=\linewidth]{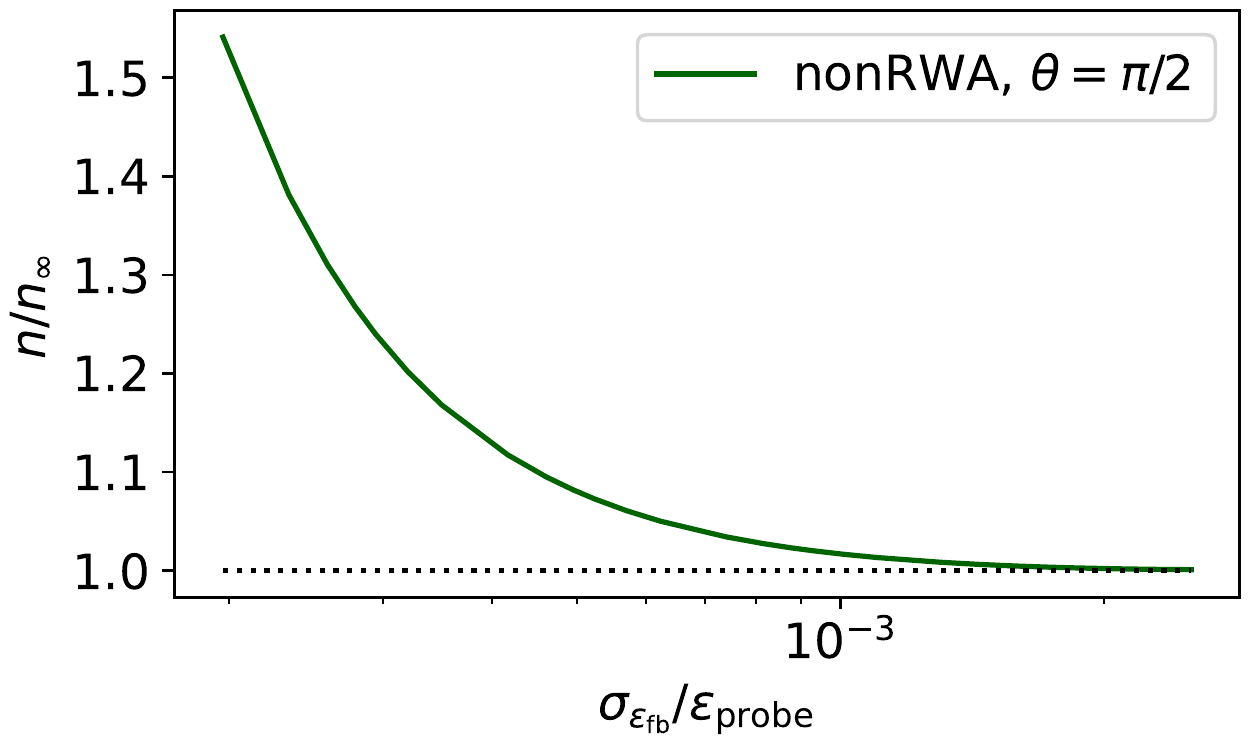}
    \caption{ Unconditional phonon number $n$ versus feedback strength $\sigma_{\epsilon_{\text{fb}}}$. The phonon number $n$ is normalized by its value at infinite feedback strength $n_{\infty} = 1.38$, while $\sigma_{\epsilon_{\text{fb}}}$ is normalized by the probe strength $\epsilon_{\text{probe}} = \SI{2.0e6}{Hz}$. The nonRWA model is used for $\theta = \pi/2$. The dotted line marks the asymptotic value $n/n_{\infty} = 1$. Feedback cost parameters range from $p/q = 10^{5}$ to $p/q = 10^{10}$. Other parameters used for this plot are $\Omega_m = 2\pi\cdot\SI{1.139}{MHz}$, $Q_m = 1.03\cdot 10^9$, $\kappa = 2\pi\cdot\SI{15,9}{MHz}$, $\eta = 0.77$, $T = \SI{300}{K}$ and $g = \SI{3.1e5}{Hz}$.}
    \label{fig:supp-nfrac-vs-fbfrac}
\end{figure}

In Fig.~\ref{fig:supp-nfrac-vs-fbfrac}, we compare the unconditional phonon number $n$ at a finite feedback strength $\sigma_{\epsilon_{\text{fb}}}$ with the phonon number at $p/q \rightarrow \infty$, denoted $n_{\infty}$. The plot is for the nonRWA model, but the results are identical to the RWA model in this case. As seen on the figure, $n$ is already very close to $n_{\infty}$ at $\sigma_{\epsilon_{\text{fb}}}/\epsilon_{\text{probe}} = 10^{-3}$. In other words, a feedback strength far less than the probe strength is needed to reach the limit unconditional phonon number. This example shows that, for at least one set of parameters, the limit unconditional states presented in the paper are approximately reachable experimentally without compromising linearity.

%



%

\end{document}